\documentclass[%
reprint,
 amsmath,amssymb,
 aps,
]{revtex4-1}

\usepackage{physics}
\usepackage{graphicx}
\usepackage{dcolumn}
\usepackage{bm}
\usepackage{float}
\usepackage[table]{xcolor}
\definecolor{Gray}{gray}{0.9}
\definecolor{LightCyan}{rgb}{0.88,1,1}
\newcolumntype{g}{>{\columncolor{Gray}}c}
\newcolumntype{q}{>{\columncolor{LightCyan}}c}

\begin{document}

\preprint{APS/123-QED}

\title{Resonant and non-resonant integrated third order parametric down-conversion}

\author{Milica Banic$^1$, Marco Liscidini$^2$,  J. E. Sipe$^1$}

\address{
$^1$ Department of Physics, University of Toronto, 60 St. George Street, Toronto, ON, M5S 1A7, Canada\\
$^2$ Department of Physics, University of Pavia, Via Bassi 6, 1-27100, Pavia, Italy
}




\date{\today}

\begin{abstract}
Third order parametric down-conversion describes a class of nonlinear processes in which a pump photon can be down-converted into triplets of photons. It has been identified as a source of non-classical light, with capabilities beyond those offered by better-established processes such as spontaneous four-wave mixing. Here we discuss the implementation of TOPDC in integrated photonic systems. We derive equations for the rates of TOPDC in a non-resonant (waveguide) and resonant (microring) platform, such that the scaling with experimental parameters can be plainly seen. We find that generally non-resonant platforms should be pursued for spontaneous TOPDC (SpTOPDC), whereas resonant platforms are more suitable for stimulated TOPDC (StTOPDC). We present a sample calculation for TOPDC rates in sample systems with conservative and accessible parameters. We find that StTOPDC should be observable with the current fabrication technology, and that with some progress in the design of TOPDC platforms, integrated SpTOPDC too could be demonstrated in the near term.
\end{abstract}

\maketitle


\section{Introduction}

Integrated optical structures, with their scalability, stability, and relative ease of fabrication, are a key tool in the development of quantum technology. Due to the interest these systems have drawn over the past decades, the efficiency of integrated platforms has improved rapidly. We are now at a stage where the generation and manipulation of non-classical light is commonplace and well understood \cite{tutorial,optimized_pairs_Harder,PhysRevLett.116.143601_squeezed_harder,zhang_nat_comm_2021}.

Parametric nonlinear processes are arguably the most common sources of non-classical light. Spontaneous four-wave mixing (SFWM) and spontaneous parametric down-conversion (SPDC), in which pairs of photons are generated by the annihilation of pump photons, are often employed as sources of photon pairs or squeezed light. One can also consider other nonlinear processes as sources of non-classical light; in this manuscript we focus on third-order parametric down-conversion (TOPDC). It consists of the down conversion of a single pump photon into a triplet of photons. This can be considered analogous to SPDC, but it relies on a third-order nonlinearity, like SFWM. Due to the use of a single pump, and the large frequency spread -- which leads to difficulties with phase matching and a relatively low mode overlap in integrated structures -- TOPDC is relatively inefficient and difficult to implement. 


Unlike SPDC and SFWM, spontaneous TOPDC could be employed as a source of non-Gaussian states or heralded photon pairs. These prospects have led to interest in TOPDC despite the difficulties in implementing it; bulk systems \cite{dot_pra_2012}, superconducting systems \cite{PhysRevX.10.011011}, and optical fibers \cite{tpg2} have been considered. {Stimulated TOPDC is also of interest; this would be easier to implement in the short term, and it could serve both as a source of photon pairs \cite{Banic_StTOPDC:22}, and as a means to study spontaneous TOPDC via spontaneous emission tomography \cite{dominguez-serna_third-order_2020}.} Surprisingly, there has been little discussion of TOPDC in integrated structures, despite the ongoing improvements in the performance of these devices. In this manuscript, we discuss the implementation of TOPDC in non-resonant and resonant optical components.

In Section \ref{section:WG} we derive expressions for the rates of spontaneous and stimulated TOPDC in a waveguide, which we take as a representative non-resonant system. In Section \ref{section:ring}, we derive the analogous expressions for TOPDC in a ring resonator, our sample resonant system. In Section \ref{section: comparison} we present a sample calculation, applying our expressions to a silicon nitride waveguide and ring resonator. We comment on the suitability of non-resonant vs resonant systems for spontaneous and stimulated TOPDC, and we identify some parameters that if improved, could lead to a viable platform for integrated TOPDC. In Section \ref{section:conclusions} we draw our conclusions. 

\section{Waveguide} \label{section:WG}

We begin by introducing the fields and the Hamiltonian in a waveguide. In our discussion of TOPDC, we will refer to a number of different frequency ranges: For example, we will distinguish the range of frequencies of the pump field, and the range of frequencies in which photons are generated. As well, we will deal with modes of different transverse spatial profiles. It is thus useful to divide the field into different ``bands" $J$, where $J$ labels both a spatial profile and a range of frequencies centered at $\omega_{J}$ with wavenumber $k_{J}$.  With this division,
the displacement field bound to a waveguide running in the $z$ direction can be written as
\begin{align}
 & \boldsymbol{D}(\boldsymbol{r})=\sum_{J}\int dk\sqrt{\frac{\hbar\omega_{Jk}}{4\pi}}a_{J}(k)\boldsymbol{d}_{Jk}(x,y)e^{ikz}dk+H.c.\label{eq:Dchannel}
\end{align}
The $a_J(k)$ are the mode operators that satisfy the usual bosonic commutation relations, $\omega_{Jk}$ is the frequency of a field component in
band $J$ at
$k$, and expanding about $k_J$ we can write 
\begin{align}
    \omega_{Jk} = \omega_J + v_J(k-k_J) + \frac{1}{2}\frac{\partial^2\omega_{Jk}}{\partial k^2}(k-k_J)^2 +... \label{eq:disp1}
\end{align}
where $v_J = \partial\omega_{Jk}/\partial k$ denotes the group velocity, and the derivatives are evaluated at $k=k_J$.   
The mode profiles $\boldsymbol{d}_{Jk}(x,y)$ are
normalized according to
\begin{align}
 & \int\int\frac{\boldsymbol{d}_{Jk}^{*}(x,y)\cdot\boldsymbol{d}_{Jk}(x,y)}{\epsilon_{0}\varepsilon_{1}(x,y;\omega_{Jk})}\frac{v_{p}(x,y;\omega_{Jk})}{v_{g}(x,y;\omega_{Jk})}dxdy=1, \label{eq:dnorm}
\end{align}
where $\varepsilon_{1}(x,y;\omega_{Jk})$ is the relative dielectric constant, and $v_{p}(x,y;\omega_{Jk})$ and $v_{g}(x,y;\omega_{Jk})$ are the local phase and group velocities, respectively. This normalization condition accounts for material dispersion in the waveguide \cite{tutorial}; the normalization in the absence of dispersion is recovered by ignoring any dependence of the quantities on $\omega_{Jk}$, and setting the local phase and group velocities equal. 

The linear Hamiltonian for the fields in the isolated waveguide is simply
\begin{align}
 & H_{L}=\sum_{J}\int dk\:\hbar\omega_{Jk}a_{J}^{\dagger}(k)a_{J}(k), \label{eq:HL}
\end{align}
and we assume that the waveguide is short enough that scattering losses can be excluded from our model. We further specialize to the case where the bands are narrow enough in frequency that to good approximation we can write $\boldsymbol{d}_{Jk}(x,y)\approx\boldsymbol{d}_{J}(x,y)$,
where the latter is $\boldsymbol{d}_{Jk}(x,y)$ evaluated at $k_{J}$,
and that in the prefactor of (\ref{eq:Dchannel}) we can approximate $\omega_{Jk}$ as the center frequency $\omega_{J}$. Under these
conditions we have 
\begin{align}
 & \boldsymbol{D}(\boldsymbol{r})=\sum_{J}\sqrt{\frac{\hbar\omega_{J}}{2}}\boldsymbol{d}_{J}(x,y)\psi_{J}(z)e^{ik_{J}z}+H.c., \label{eq:Dchannel2}
\end{align}
where 
\begin{align}
 & \psi_{J}(z)=\int\frac{dk}{\sqrt{2\pi}}a_{J}(k)e^{i(k-k_{J})z}.\label{eq:field_operator}
\end{align}
Using (\ref{eq:HL}) and (\ref{eq:field_operator}), the linear Hamiltonian can be rewritten in terms of the field operators $\psi_{J}(z)$. Using the dispersion relation (\ref{eq:disp1}) we find
\begin{align}
\nonumber
 & H_{L}=\sum_{J}\hbar\omega_{J}\int\psi_J^{\dagger}(z)\psi_J(z)dz\\
 &-\frac{1}{2}i\hbar v_{J}\int\left(\psi_{J}^{\dagger}(z)\frac{\partial\psi_{J}(z)}{\partial z}-\frac{\partial\psi_{J}^{\dagger}(z)}{\partial z}\psi_{J}(z)\right)dz + ..., \label{eq:HL_psi}
\end{align}
where the ellipses denote higher order dispersion terms 
. 

Having established the waveguide's linear Hamiltonian, we now address the nonlinear behaviour. For a $\chi_{3}$ nonlinearity, the nonlinear Hamiltonian is
\begin{align}
 H_{NL}=\nonumber -\frac{1}{4\epsilon_{0}}\int\Gamma^{ijkl}(\boldsymbol{r};\{\omega_J\})&D_{J1}^{i}(\boldsymbol{r})D_{J2}^{j}(\boldsymbol{r})D_{J3}^{k}(\boldsymbol{r})\\ &\times D_{J4}^{l}(\boldsymbol{r})d\boldsymbol{r}\label{eq:HNL},
\end{align}
where the subscripts label the different modes involved in the interaction, and the lowercase superscripts denote Cartesian components of the field. We use $\{\omega_J\}$ to denote the frequencies of the four modes involved in the nonlinear interaction; that is, $\Gamma^{ijkl}(\boldsymbol{r};\{\omega_J\})$ should be read as $\Gamma^{ijkl}(\boldsymbol{r};\omega_{J1},\omega_{J2},\omega_{J3},\omega_{J4})$. We take
\begin{align}
    \Gamma_{3}^{ijkl}(\bold{r};\{\omega_{J}\}) = \frac{\chi_{3}^{ijkl}(\bold{r};\{\omega_{J}\}) }{\epsilon_0^2 \varepsilon_{1}(\bold{r};\omega_{J1})\varepsilon_{1}(\bold{r};\omega_{J2})\varepsilon_{1}(\bold{r};\omega_{J3})\varepsilon_{1}(\bold{r};\omega_{J4})},
\end{align}
where we have neglected any contribution to $\Gamma_{3}^{ijkl}(\bold{r};\{\omega_{J}\})$ due to any $\chi^{(2)}$ in the system; this contribution could be included as done in other work \cite{tutorial}. We introduce a general nonlinear parameter
\begin{align}
    \gamma_{J1,J2,J3,J4} = \frac{3(\omega_{J1}\omega_{J2}\omega_{J3}\omega_{J4})^{1/4}\bar{\chi}_{3}}{4\epsilon_0(\bar{n}_{J1}\bar{n}_{J2}\bar{n}_{J3}\bar{n}_{J4})^{1/2}c^2} \frac{e^{i\Phi_{J1,J2,J3,J4}}}{\mathcal{A}_{J1,J2,J3,J4}}, \label{eq:general_gamma}
\end{align}
where $\bar{\chi}_{3}$ and $\bar{n}$ {are nominal values of $\chi_{3}^{ijkl}(\bold{r};\{\omega_{J}\})$ and the index of refraction, which have spatial and frequency dependence in general}. The parameter $\mathcal{A}_{J1,J2,J3,J4}$ is an effective area for the process. Its magnitude is determined by the overlap between the interacting modes and the material's nonlinearity; the precise definition of $\mathcal{A}_{J1,J2,J3,J4}$ is given in Appendix \ref{appendix:a_eff}. The factor $e^{i\Phi_{J1,J2,J3,J4}}$ is introduced to ensure that the effective area is a real number, since the mode overlap is in general complex. 

\subsection{Self- and cross-phase modulation}
\label{section:WG_PM}

With currently available integrated photonic structures, a practical TOPDC system will require the use of a strong pump field due the weak nonlinearity. The self- and cross-phase modulation (SPM and XPM) due to the pump field cannot be neglected; here we discuss these effects in the waveguide. We assume that any stimulating fields are weak compared to the pump, such that phase modulation effects due to their presence are negligible. 

We start by considering SPM on the pump field. Into Eq.  (\ref{eq:HNL}) we substitute (\ref{eq:Dchannel}), where the displacement field has a term associated with each of the four modes, namely the pump which we label with $P$, and the three generated modes $G_1$, $G_2$, and $G_3$. Expanding the displacement fields in (\ref{eq:HNL}) and collecting the terms responsible for SPM on the pump, we find
\begin{align}
 & H_{SPM}=-\gamma_{SPM}\frac{\hbar^{2}\omega_{P}v_{P}^{2}}{2}\int dz\psi_{P}^{\dagger}(z)\psi_{P}^{\dagger}(z)\psi_{P}(z)\psi_{P}(z),
\end{align}
where $\gamma_{SPM}$ is defined in the usual way \cite{tutorial}, such that $\gamma_{SPM} = \gamma_{PPPP}$ with the latter defined according to \eqref{eq:general_gamma}. 
To obtain the nonlinear phase shift associated with SPM, we consider the evolution of the pump field operator $\psi_P(z,t)$ due to $H_L + H_{SPM}$. In the Heisenberg picture we have
\begin{align}
\nonumber
 & i\hbar\frac{\partial\psi_{P}(z,t)}{\partial t}=\hbar\omega_{P}\psi_{P}(z,t)-i\hbar v_{P}\frac{\partial\psi_{P}(z,t)}{\partial z}\\
 & -\gamma_{SPM}\hbar^{2}\omega_{P}v_{P}^{2}\psi_{P}^{\dagger}(z,t)\psi_{P}(z,t)\psi_{P}(z,t), \label{eq:HeisenbergSPM}
\end{align}
neglecting group velocity dispersion and higher order dispersion terms in $H_L$. Then taking the classical limit for this strong field ($\psi_{P}(z,t)\rightarrow\phi_{P}(z,t)$), and seeking a solution with the form $\phi(z,t)=\phi(z)\exp(-i\omega_{P}t)$, we find
\begin{align}
& \frac{\partial\phi_{P}(z)}{\partial z}=i\gamma_{SPM}P_{P}(z)\phi_{P}(z),
\end{align}
where
\begin{align}
& P_{P}(z)=\hbar\omega_{P}v_{P}\left|\phi_{P}(z)\right|^{2}
\end{align}
is the power in the pump field \cite{tutorial}. For a CW pump, this is constant over $z$ and the solution in the nonlinear region is
\begin{align}
 & \phi(z,t)={\phi}_{P}e^{i\gamma_{SPM} P_P (z+L/2)}e^{-i\omega_{P}t}\label{eq:NLform},
\end{align}
where we have taken the nonlinear region to span from $z=-L/2$ to $z=L/2$. 
Putting (\ref{eq:NLform}) into (\ref{eq:Dchannel2}), we find the pump field under $H_L+H_{SPM}$ to be
\begin{align}
 & \boldsymbol{D}_{P}(\boldsymbol{r},t)=\sqrt{\frac{\hbar\omega_{P}}{2}}\boldsymbol{d}_{P}(x,y){\bar{\phi}}_{P}e^{i\bar{k}_{P}z}e^{-i\omega_{P}t}+c.c.\label{eq:DT_SPM},
\end{align}
where $\bar{k}_{P}=k_{P}+\gamma_{SPM}P_{P}$, and $\bar{\phi}_P = \phi_P e^{i(\bar{k}_P - k_P)L/2}$. One can see that the effect of SPM on the pump is to take $k_P \rightarrow \bar{k}_P$; we have recovered the expected nonlinear phase shift, which is set by the pump power $P_P$ and the nonlinear parameter $\gamma_{SPM}$.

We apply a similar approach to account for the generated modes' phase shifts due to XPM by the pump. We again expand (\ref{eq:HNL}) using the full displacement field, now collecting terms associated with XPM of mode $G$ by the pump. Taking the pump to be a classical, CW field, we have 
\begin{align}
    & H_{XPM}=-2\gamma_{XPM}P_{P}\hbar v_{G}\int\psi_{G}^{\dagger}(z)\psi_{G}(z)dz,
\end{align}
where $\gamma_{XPM}$ is defined in the usual way \cite{tutorial}; we have
\begin{align}
\gamma_{XPM} = \sqrt{\frac{\omega_G}{\omega_P}} \gamma_{PGPG}, \label{eq:gamma_XPM} 
\end{align}
with $\gamma_{PGPG}$ defined according to \eqref{eq:general_gamma}.

When treating XPM in modes where we consider the generation of photons, we must account for dispersion, since photons can be generated over a range of wavenumbers. To do this, we keep higher order terms in (\ref{eq:HL_psi}); the equation of motion for field $G$ under $H_L + H_{XPM}$ is then 
\begin{align}
\nonumber
 & i\hbar\frac{\partial\psi_{G}(z,t)}{\partial t}= -2 \gamma_{XPM} P_P \hbar v_G \psi_{G}(z,t) + \hbar\omega_{G}\psi_{G}(z,t)\\ &-i\hbar v_{G}\frac{\partial\psi_{G}(z,t)}{\partial z} + ... \label{eq:HeisenbergXPM_full}
\end{align}
where the ellipses denote terms arising from terms of the second order and higher in the dispersion relation (\ref{eq:disp1}). In the absence of a nonlinearity, the solution to the Heisenberg equation has the form 
\begin{align}
    \psi_G(z,t) = \int \frac{dk}{\sqrt{2\pi}} a_G(k) e^{i(k-k_G)z}e^{-i\omega_{Gk}t}. \label{eq:psiG_outside}
\end{align}
Inside the nonlinear region, which spans from $z=-L/2$ to $z=L/2$, we seek a solution of the form
\begin{align}
    \psi_G(z,t) = \int \frac{dk}{\sqrt{2\pi}} \bar{a}_G(k) e^{i(\bar{k}-k_G)z}e^{-i\omega_{Gk}t}, \label{eq:psiG_inside}
\end{align}
{where $\bar{a}_G(k)$ and $\bar{k}$ are to be determined}. The displacement field in the nonlinear medium is then
\begin{align}
 & \boldsymbol{D}_{G}(\boldsymbol{r},t)=\sqrt{\frac{\hbar\omega_{G}}{4\pi}}\boldsymbol{d}_{G}(x,y)\int dk\:\bar{a}_{G}(k)e^{i\bar{k}z}e^{-i\omega_{Gk}t}+H.c. \label{eq:DG_XPM}
\end{align}

Taking $\psi_G(z,t)$ for $z>L/2$ as defined by (\ref{eq:psiG_outside}) and imposing continuity for the field at $z=L/2$, we immediately see that $\bar{a}_G(k) = a_G(k)e^{i(k-\bar{k})L/2}$. Putting (\ref{eq:psiG_inside}) into the full Heisenberg equation  (\ref{eq:HeisenbergXPM_full}), we have
\begin{align}
    \omega_{Gk} = - 2 \gamma_{XPM} v_G P_P + \omega_G + \frac{\partial \omega_{Gk}}{\partial \bar{k}} (\bar{k} - k_G) + ..., \label{eq:EOM_solution}
\end{align}
where again the ellipses denote higher order derivatives, similar to those in (\ref{eq:disp1}). Taking this with (\ref{eq:disp1}), and neglecting {what we find to be} the small difference between ${\partial\omega_{Gk}}/{\partial \bar{k}}$ and ${\partial\omega_{Gk}}/{\partial {k}}$, we have 
\begin{align}
    \bar{k} 
    &= k + 2 \gamma_{XPM} P_P + ...
    \label{eq:SPM_shift}
\end{align}
Introducing the shifted reference wavevector $\bar{k}_G \equiv \bar{k}(\omega_G)$, we also have 
\begin{align}
    \bar{k}_G
    &= k_G + 2 \gamma_{XPM} P_P + ...
\end{align}
With this we have specified the form of the generated fields under $H_L + H_{SPM} + H_{XPM}$. 

When discussing StTOPDC configurations, we will neglect the generation of photons in the seeded mode $G \equiv S$, instead simply treating these modes as classical. Furthermore, we will work with CW fields such that dispersion terms in the general expression (\ref{eq:DG_XPM}) can be ignored. Eq. \eqref{eq:DG_XPM} then reduces to
\begin{align}
 & \boldsymbol{D}_{S}(\boldsymbol{r},t)=\sqrt{\frac{\hbar\omega_{S}}{2}}\boldsymbol{d}_{S}(x,y){\bar{\phi}}_{S}e^{i\bar{k}_{S}z}e^{-i\omega_{S}t}+c.c.\label{eq:DS_XPM},
\end{align}
with $\bar{k_S} = k_S + 2 \gamma_{XPM} P_P$ and $\bar{\phi}_S = \phi_S e^{i(\bar{k}_S - k_S)L/2}$. Here again we recover the expected phase shift due to XPM on the seed.

{Having treated the effects of SPM and XPM in the steady state limit, and we can now study TOPDC for a CW pump with these effects taken into account. A more careful treatment would be required to model account for SPM and XPM in pulsed-pump TOPDC, but we defer this problem to future work, specializing here to the CW case.}

\subsection{Non-resonant triplet generation}\label{section:TOPDC_WG}

We now begin our discussion of TOPDC efficiency in a waveguide. TOPDC consists of the downconversion of a pump photon into a triplet of photons in modes $G_1$, $G_2$, and $G_3$; we label the interaction Hamiltonian describing this $H_{G_1 G_2 G_3}$. We adopt an interaction picture such that the full Hamiltonian is split up as $H = H_0 + H_{G_1 G_2 G_3}$, where $H_0 = H_L + H_{SPM} + H_{XPM}$. Then the 
ket $\left|\Psi(t)\right\rangle$ evolves according to 
\begin{align}
 & i\hbar\frac{d}{dt}\left|\Psi(t)\right\rangle =H_{G_1 G_2 G_3}^{(I)}(t)\left|\Psi(t)\right\rangle ,\label{eq:Schroedinger}
\end{align}
and $H_{G_1 G_2 G_3}^{I}(t) = e^{i H_0 t/\hbar}H_{G_1 G_2 G_3}e^{-i H_0 t/\hbar}$. Similarly to Section \ref{section:WG_PM}, we write $H_{G_1 G_2 G_3}$ by expanding (\ref{eq:HNL}) with the full waveguide displacement fields, and here collecting terms involving one lowering operator for the pump, and three raising operators for the generated modes. To impose the interaction picture, we simply expand (\ref{eq:HNL}) in the time-dependent displacement fields that evolve under $H_0$; for a CW pump, these are Eqs. (\ref{eq:DT_SPM}), (\ref{eq:DS_XPM}), and (\ref{eq:DG_XPM}).  

The iterative solution of (\ref{eq:Schroedinger}) 
is 
\begin{align}
 & \left|\Psi(t)\right\rangle =\left|vac\right\rangle - \frac{i}{\hbar}\int_{-\frac{T}{2}}^{\frac{T}{2}}H_{G_1 G_2 G_3}^{(I)}(t)\left|\text{vac}\right\rangle dt+...,\label{eq:psi_evolve}
\end{align}
so to first order, the probability of a TOPDC interaction occurring within a time $T$ is  
\begin{align}
 & \mathcal{P}=\frac{1}{\hbar^{2}}\int_{-\frac{T}{2}}^{\frac{T}{2}}dt\int_{-\frac{T}{2}}^{\frac{T}{2}}dt'\left\langle \text{vac}|H_{G_1 G_2 G_3}^{(I)}(t')H_{G_1 G_2 G_3}^{(I)}(t)|\text{vac}\right\rangle, \label{eq:Pwork}
\end{align}
and the interaction rate is ${R}_{\text{int}} = \mathcal{P}/T$. We will also refer to the efficiency $R_{\text{int}}/R_P$, where $R_P$ is the rate of incoming pump photons.

\subsubsection{Spontaneous triplet generation}

\begin{figure}[h]
    \centering
    \includegraphics[width=0.48\textwidth]{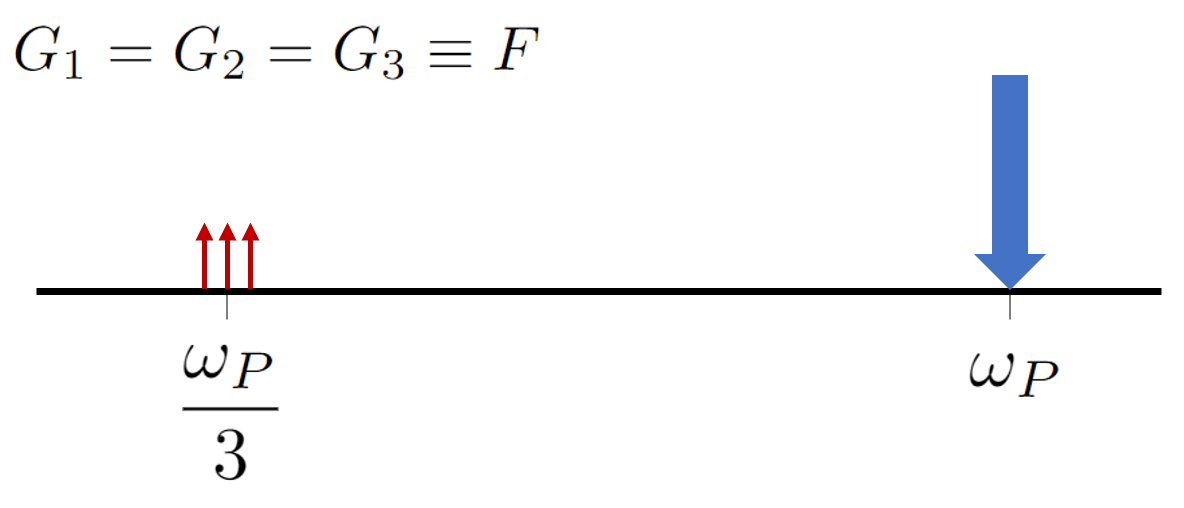}
    \caption{Mode configuration for degenerate spontaneous TOPDC. Arrows pointing up represent generated modes; the arrow pointing down represents an input mode. {The range over which photons can be generated around $\omega_P/3$ depends on the system; for example, the range will be tighter in a ring resonator than in a waveguide due to the narrow resonant linewidth.}}
    \label{fig:degenSpTOPDC}
\end{figure}

We begin by discussing spontaneous TOPDC (SpTOPDC). We consider the degenerate configuration in which the three photons are generated in the same mode ($G_1 = G_2 = G_3$), from a CW pump at the third harmonic frequency. We expand the general nonlinear Hamiltonian (\ref{eq:HNL}) with the fields (\ref{eq:DT_SPM}) and (\ref{eq:DG_XPM}). Collecting TOPDC terms where $G_1 = G_2 = G_3 \equiv F$, we find the interaction Hamiltonian

\begin{align}
\nonumber
 & H_{FFF}^{(I)}(t)=-\int dk_{1}dk_{2}dk_{3}M_{FFF}(k_{1},k_{2},k_{3})\bar{a}_{F}^{\dagger}(k_{1})\\
 &\times \bar{a}_{F}^{\dagger}(k_{2})\bar{a}_{F}^{\dagger}(k_{3})e^{-i\Omega_{FFF}(k_{1},k_{2},k_{3})t}+H.c.,\label{eq:HFFF}
\end{align}
where 
\begin{align}
 & \Omega_{FFF}(k_{1},k_{2},k_{3})=\omega_{P}-\omega_{Fk_{1}}-\omega_{Fk_{2}}-\omega_{Fk_{3}},
 \label{eq:Omega_FFF}
\end{align}
and 
\begin{align}
\nonumber
 & M_{FFF}(k_{1},k_{2},k_{3})= \hbar^2 L \frac{\omega_F}{3 \sqrt{2\pi}^3} \bar{\phi}_{P} \sqrt{v_T v_F^3} \gamma_{FFF}\\ &\times \text{sinc}\left(\frac{\Delta\bar{k}_{FFF} L}{2}\right) \label{eq:MFFF}.
\end{align}
Here 
\begin{align}
    \Delta \bar{k}_{FFF} = \bar{k}_{P}-\bar{k}_{1}-\bar{k}_{2}-\bar{k}_{3} \label{eq:mismatch_Sp}
\end{align}
is the wavenumber mismatch, and 
\begin{align}
    \gamma_{FFF} = \frac{(\omega_F^3 \omega_P)^{1/4}}{\omega_F} \gamma_{FFFP}, \label{eq:gamma_FFF}
\end{align}
where $\gamma_{FFFP}$ is defined according to \eqref{eq:general_gamma}.


Using Eq. (\ref{eq:HFFF}) with (\ref{eq:Pwork}), we find 
\begin{align}
    {R}_{FFF}/R_P=(|\gamma_{FFF}| L)^{2}P_{\text{vac}}^2, \label{eq:RFFF_WG}
\end{align} 
where we have introduced a characteristic vacuum power; that is, the power associated with the quantum fluctuations in the generated mode \cite{how_does_it_scale}. This can be generally defined as 
\begin{align}
    P_{\text{vac}} = \frac{\hbar \bar{\omega}}{\tau}, \label{eq:Pvac}
\end{align}
where $\bar{\omega}$ is a characteristic frequency set by the product of the generated modes' frequencies. For example, in SpTOPDC we have $\bar{\omega} = (\omega_{G_1}\omega_{G_2}\omega_{G_3})^{1/3}$, which reduces to $\bar{\omega} = \omega_F$ in this degenerate configuration. By $\tau^{-1}$ we denote the frequency bandwidth over which the photons can be generated. This generation bandwidth depends both on the nonlinear process and structure under consideration, therefore each case discussed in this work will have a distinct $\tau^{-1}$. For SpTOPDC in the waveguide, we find the generation bandwidth to be
\begin{align}
 \tau^{-2}_{FFF} = \frac{v_{F}^{3}}{6\pi^{2}} \nonumber
 \int & dk_{1}dk_{2}dk_{3}\delta(\Omega_{FFF}(k_1,k_2,k_3))\\ &\times \text{sinc}^{2}\left(\frac{\Delta\bar{k}_{FFF}{L}}{2}\right). \label{eq:WG_bw2}
\end{align}

From Eq. \eqref{eq:WG_bw2} we see that the generation bandwidth is determined by integrating over the values that the generated photons' wavenumbers can take, while satisfying energy conservation and phase matching. The former condition is imposed in (\ref{eq:WG_bw2}) by the Dirac delta function, the latter by the sinc function. The bandwidth over which these conditions are simultaneously satisfied depends on the waveguide's dispersion properties. 
In this case there are three free wavenumbers, corresponding to the three photons generated, and the bandwidth over which these wavenumbers can vary determines $\tau^{-2}$ due to the generation rate's quadratic scaling with vacuum power. Since the vacuum power scales with the generation bandwidth, a higher bandwidth can lead to a higher generation rate in the waveguide. The relationship between the generation bandwidth and the generation rate is nuanced; for example, if the length of the waveguide is increased, we will see that the generation bandwidth decreases, while the overall generation rate increases because of the quadratic scaling with $L$ in Eq. \eqref{eq:RFFF_WG}.

One can consider a simple limit in which (\ref{eq:WG_bw2}) can be evaluated analytically, as described in Appendix \ref{section:BW}. Assuming that higher order dispersion terms are weak, such that we can work to second order in (\ref{eq:disp1}), and assuming phase matching ($\bar{k}_P - 3\bar{k}_F = 0$), we find
\begin{align}
    \tau^{-2}_{FFF} = \frac{\sqrt{3}}{9|\beta_2|L}. \label{eq:analytictau3PG}
\end{align}
In this limit, the generation bandwidth is set by $L$ and {the group velocity dispersion $\beta_2 = \frac{\partial^2 k}{\partial \omega_{Jk}^2}$}. In general, higher-order dispersion terms may be important and the generation bandwidth must be computed numerically. 

\subsubsection{Stimulated triplet generation}\label{Section:StTOPDC_WG}

\begin{figure}[h]
    \centering
    \includegraphics[width=0.5\textwidth]{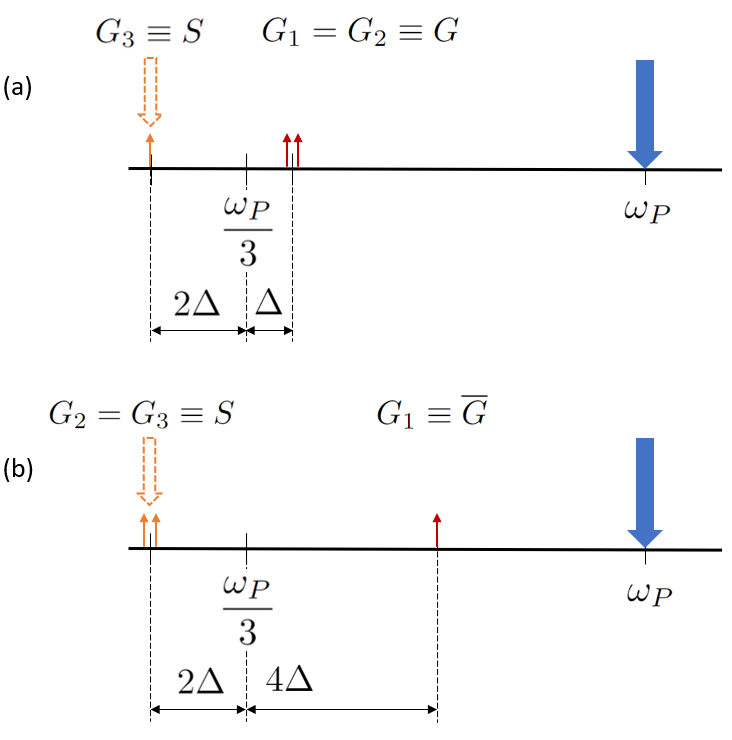}\caption{Sketch of the mode configurations considered in our discussion of a) StTOPDC and b) DStTOPDC. Arrows pointing up represent generated modes; arrows pointing down represent input modes. The solid and dashed arrows represent pump and seed fields, respectively. {In StTOPDC the range over which pairs can be generated around depends on the system; in a resonator the photons' frequencies are tightly restricted due to the narrow resonances, whereas in a waveguide the frequencies at which photons are generated can vary more.}}
    \label{fig:StTOPDC}
\end{figure}

We now discuss stimulated TOPDC (StTOPDC), where we seed the process with a classical CW field at $\omega_S$. In our treatment, we neglect the description of stimulated photons at $\omega_S$. One case that arises then is the `singly stimulated' case, where one photon is generated at $\omega_S$ with the other two centered at $\omega_G$, such that $\omega_P = \omega_S + 2\omega_G$, as indicated in Fig. \ref{fig:StTOPDC}a. 
We will also consider the `doubly stimulated' case in which two photons are generated at $\omega_S$ with one photon at $\omega_G$, giving $\omega_P = 2\omega_S + \omega_G$, sketched in Fig. \ref{fig:StTOPDC}b. The latter process is in general more efficient, due to the Hamiltonian's higher scaling with the seed field's amplitude. However, one could suppress this process by tuning $\omega_S$; for example, setting $\omega_S = \omega_P/2$ would make the latter energy conservation condition impossible to satisfy, allowing only the singly stimulated process. 

We first address the singly stimulated case sketched in Fig. \ref{fig:StTOPDC}a, again constructing the interaction Hamiltonian by collecting the relevant subset of terms in (\ref{eq:HNL}); namely, we keep TOPDC terms with $G_1 = G_2 \equiv G$, $G_3 \equiv S$. We treat the seeded mode classically by taking $\bar{a}^{\dagger}_S(k) \rightarrow \alpha^*_S(k)$. For a CW seed we have the interaction Hamiltonian
\begin{align}   
    \nonumber
 & H_{GG(S)}^{(I)}(t)=-\int dk_{1}dk_{2}M_{GG(S)}(k_{1},k_{2})\bar{a}_{G}^{\dagger}(k_{1})\bar{a}_{G}^{\dagger}(k_{2})\\\
 &\times e^{-i\Omega_{GG(S)}(k_{1},k_{2})t}+H.c.\label{eq:HGG(S)}
\end{align}
Here we have defined
\begin{align}
 & \Omega_{GG(S)}(k_1,k_2)=\omega_{P}-\omega_{S}-\omega_{Gk_{1}}-\omega_{Gk_{2}},
\end{align}
and
\begin{align}
 & \nonumber M_{GG(S)}(k_{1},k_{2})= L \hbar^2 \sqrt{\omega_G \omega_S}(\omega_{P}\omega_{S}\omega_{G}^2)^{1/4} \frac{1}{2\pi} \bar{\phi}_{P}\bar{\phi}_{S}^{*}
 \\
 &\times \sqrt{v_{P}v_{S}v_{G}^2} \gamma_{GG(S)} \: \text{sinc}\left(\frac{\Delta\bar{k}_{GG(S)} L}{2}\right), 
\end{align}
with $\Delta \bar{k}_{GG(S)} = \bar{k}_{P}-\bar{k}_{S}-\bar{k}_{1}-\bar{k}_{2}$, and 
\begin{align}
    \gamma_{GG(S)} =\left( \frac{\omega_P}{\omega_S}\right)^{1/4} \gamma_{GGSP} \label{eq:gamma_GGS},
\end{align}
again defining $\gamma_{GGSP}$ according to \eqref{eq:general_gamma}.

We now have an effective pair generation Hamiltonian, due to our classical treatment of the seeded mode. It should be emphasized that although our model neglects the generation of photons in the seeded mode, \eqref{eq:HGG(S)} still describes a TOPDC process. The energy conservation condition is unchanged, and the interaction rate computed for this Hamiltonian is in principle a triplet generation rate; in our treatment we effectively trace over the mode $S$, excluding the photons generated there. In mode $G$ where two of three photons are emitted, we find the normalized interaction rate
\begin{align}
 & {R}_{GG(S)}/R_P=(\gamma_{GG(S)}L)^2 P_S P_{\text{vac}}, \label{eq:RS3WG}
\end{align}
where $P_{\text{vac}}$ for this process is
\begin{align}
    P_{\text{vac}} =  \frac{\hbar \omega_G}{\tau},
\end{align}
with the generation bandwidth
\begin{align}
 \nonumber \tau_{GG(S)}^{-1} = \frac{v_{G}^{2}}{\pi} \int & dk_{1}dk_{2}\delta(\Omega_{GG(S)}(k_1,k_2))\\
 &\times \text{sinc}^{2}\left(\frac{\Delta \bar{k}_{GG(S)} L}{2}\right) \label{eq:WG_bw}.
\end{align}
This is interpreted in the same way as Eq. (\ref{eq:WG_bw2}); it is the bandwidth over which the two photons can be generated, as constrained by energy and momentum conservation. In this case, there are two free wavenumbers, corresponding to the two photons generated. This can be computed analytically if we assume phase matching ($\bar{k}_P - \bar{k}_S - 2\bar{k}_G$) and keep terms up to third order in Eq. \eqref{eq:disp1}; a detailed discussion of this case is given in Appendix \ref{appendix:evaluatingtau}. In this limit, Eq. \eqref{eq:WG_bw} evaluates to %
\begin{align}
    \tau_{GG(S)} = \frac{4}{3}\sqrt{\frac{2}{\pi|\beta_2|L}}, \label{eq:analytictauS3}
\end{align}
which has the same scaling with $\beta_2$ and $L$ as (\ref{eq:analytictau3PG}). The discrepancy between the constant factors in (\ref{eq:analytictauS3}) and (\ref{eq:analytictau3PG}) is not surprising. While both expressions define the bandwidth over which the generated photons' wavenumbers and frequencies can vary, they are fundamentally different in that $\tau_{GG(S)}^{-1}$ is defined by considering the possible values that $k$ can take on in a pair of photons, whereas $\tau_{FFF}^{-1}$ is defined with respect to a triplet of photons. It is therefore reasonable to expect the same scaling with the waveguide parameters in both cases, but there is no reason to expect $\tau_{FFF}^{-1} = \tau_{GG(S)}^{-1}$. 

Comparing Eqs. (\ref{eq:RFFF_WG}) and (\ref{eq:RS3WG}), one can see that the vacuum power and classical seed power play analogous roles in the spontaneous and stimulated processes, respectively. Indeed, the vacuum power is defined as it is so the improvement to the efficiency with a stimulating field can be estimated by comparing the vacuum powers and seed power \cite{how_does_it_scale}. 
%
%
%
The generation rate improves by a factor proportional to $P_S/P_{\text{vac}}$ in the presence of a seed.

To illustrate this further, we turn to the `doubly stimulated' TOPDC (DStTOPDC) configuration, sketched in Fig. \ref{fig:StTOPDC}b, in which photons are generated in mode $\bar{G}$. {This process can be considered as classical, in the sense that it is not driven by vacuum fluctuations and can be described in the framework of a classical electromagnetic theory. However, DStTOPDC can also be studied within a quantum theory, as we do here for a better comparison with SpTOPDC and StTOPDC.} Treating the seeded mode classically, for a CW seed we have the Hamiltonian
\begin{align}
 & H_{\bar{G}(SS)}^{(I)}(t)=-\int dk M_{\bar{G}(SS)}(k)\bar{a}_{\bar{G}}^{\dagger}(k)e^{-i\Omega_{\bar{G}(SS)}(k)t}+H.c.,\label{eq:HDS3}
\end{align}
where 
\begin{align}
 & \Omega_{\bar{G}(SS)}(k)=\omega_{P}-2\omega_{S}-\omega_{\bar{G}k},
\end{align}
and 
\begin{align}
\nonumber & M_{\bar{G}(SS)}(k)= L \hbar^2 \omega_S {\frac{1}{\sqrt{2\pi}}}\bar{\phi}_{P}\left(\bar{\phi}_{S}^{*}\right)^{2}\sqrt{v_{P}v_{\bar{G}}v_{S}^2}\\
 & \times \gamma_{\bar{G}(SS)} \: \text{sinc}\left(\frac{\Delta \bar{k}_{\bar{G}(SS)} L}{2}\right).
\end{align}
The phase mismatch is given by $\Delta \bar{k}_{\bar{G}(SS)} = \bar{k}_{P}-\bar{2k}_{S}-\bar{k}$, and we define the nonlinear parameter
\begin{align}
    \gamma_{\bar{G}(SS)} = \left( \frac{\omega_P \omega_{\bar{G}}}{\omega_S^2}\right)^{1/4}\gamma_{\bar{G}SSP}, \label{eq:gamma_G(SS)}
\end{align}
where $\gamma_{\bar{G}SSP}$ is defined according to \eqref{eq:general_gamma}. With Eq. (\ref{eq:Pwork}), we find 
\begin{align}
 \nonumber {R}_{\bar{G}(SS)}/R_P&= \frac{1}{2\pi}(|\gamma_{\bar{G}(SS)}| L P_{S})^{2}\\ &\times \text{sinc}^{2}\left( \left(\bar{k}_{P}-2\bar{k}_{S}-\bar{k}_{\bar{G}}\right)\frac{L}{2} \right). \label{eq:RDS3WG}
\end{align}

As expected for a classical nonlinear process \cite{how_does_it_scale,Boyd2008-eu}, the efficiency of DStTOPDC scales only with the power of the seed field. Comparing (\ref{eq:RDS3WG}) to (\ref{eq:RS3WG}), we see the same correspondence between the {seeded and unseeded} processes discussed above when comparing (\ref{eq:RFFF_WG}) and (\ref{eq:RS3WG}): {The role played by the seed power in the stimulated process can be ascribed to the `vacuum power' in the corresponding unseeded process, which is driven by vacuum fluctuations and requires a quantum description}. The TOPDC generation rate in all three cases simply scales with a system dependent nonlinear parameter, $(\gamma L)^2$, and the power of each classical or quantum field driving the process. The TOPDC efficiency then improves by a factor proportional to $P_S/P_{\text{vac}}$ with each `order' of seeding. This is related to earlier work, in which spontaneous pair generation processes and their seeded counterparts were linked in the same way \cite{how_does_it_scale}. In this case we can consider two distinct `levels' of seeding, due to the fact the spontaneous process results in the generation of photon triplets rather than pairs. 

\section{Ring} \label{section:ring}

\begin{figure}
    \centering
    \includegraphics[width=0.48\textwidth]{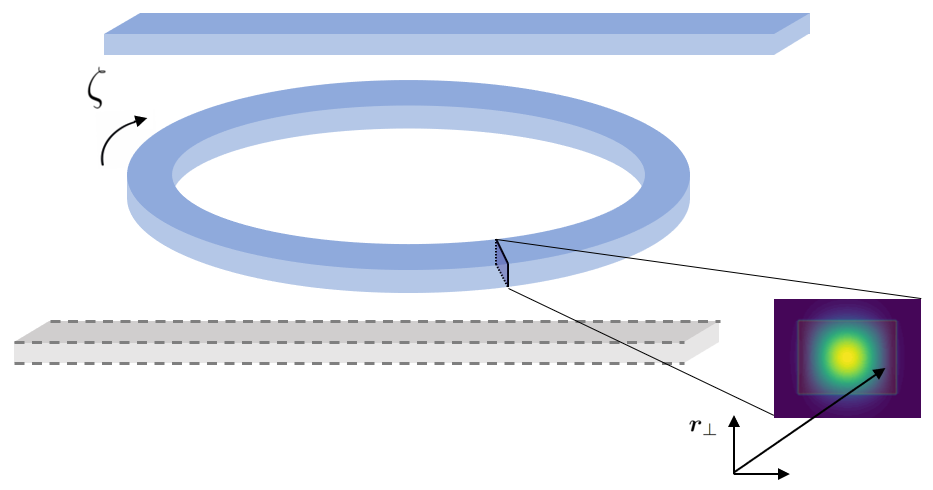}
    \caption{Sketch of the mode profile and coordinate system for a microring resonator coupled to a bus waveguide (solid blue) and a phantom waveguide (dashed grey).}
    \label{fig:ring}
\end{figure}

Having derived the efficiencies for SpTOPDC, StTOPDC, and DStTOPDC in a waveguide, we will now carry out the analogous calculation for a resonant system. We choose as our sample resonant structure a microring resonator coupled to a bus waveguide, as sketched in Fig. \ref{fig:ring}. Ring resonators are ubiquitous in integrated nonlinear optics; they are relatively easy to implement, and their resonant enhancement enables the realization of nonlinear processes with relatively high efficiency \cite{Azzini:12}. The linear dynamics of the ring-channel system are modeled by the Hamiltonian 
\begin{align}
    \nonumber
    H_L = &H_\text{ring} + H_\text{channel} + H_\text{phantom\:channel}\\ &+ H_\text{coupling} + H_\text{phantom\:coupling}, \label{eq:ring_HL}
\end{align}
where $H_\text{ring}$ and $H_\text{channel}$ capture the free evolution of fields in the ring and channel; we have
\begin{align}
    &H_\text{ring} = \sum_J \hbar \omega_J b^{\dagger}_J b_J,
\end{align}
and $H_\text{channel}$ is given in Eq. \eqref{eq:HL_psi}. We adopt a point coupling model so that
\begin{align}
    H_\text{coupling} &= \sum_J \hbar \gamma_J b^{\dagger}_J \psi_J(0) + H.c., \label{eq:H_coupling}
\end{align}
where $\gamma_J$ is a complex coupling constant \cite{Vernon_lossy_resonators,tutorial}. The `phantom channel' terms in Eq. (\ref{eq:ring_HL}) account for scattering losses from the ring \cite{Vernon_lossy_resonators}; unlike in the waveguide, scattering losses here cannot be taken to be negligible because photons dwell in the ring, rather than passing straight through the system. The phantom channel terms have the same form as Eqs. (\ref{eq:HL_psi}) and (\ref{eq:H_coupling}) with a distinct field operator
\begin{align}
 & \phi_{J}(z)=\int\frac{dk}{\sqrt{2\pi}}c_{J}(k)e^{i(k-k_{J})z}, 
\end{align}
where the $c_J(k)$ are bosonic annihilation operators. We label the phantom channel group velocities $u_J$ and the ring-phantom channel coupling constants $\mu_J$. With the coupling to the physical and phantom channels, we have a total decay constant for the ring given by 
\begin{align}
    \bar{\Gamma}_J = \frac{|\gamma_J|^2}{2 v_J} + \frac{|\mu_J|^2}{2 u_J},
\end{align}
which sets the loaded quality factor $Q_J = \omega_J/2\bar{\Gamma}_J$.

The fields in the bus waveguide are again given by Eq. (\ref{eq:Dchannel}). Inside the ring, we take the fields to consist of discrete modes, with the form
\begin{align}
    \bold{D(r},t) = \sum_{J} \sqrt{\frac{\hbar \omega_J}{2}} \frac{\bold{d}_J(\bold{r_\perp},\zeta)}{\sqrt{\mathcal{L}}} b_J(t) e^{i\kappa_J \zeta} + H.c.
    \label{eq:Dring}
\end{align}
Each mode $J$ is associated with a ring resonance, with the resonant frequency $\omega_J$ and resonant wavenumber $\kappa_J$. As sketched in Fig. \ref{fig:ring}, we adopt a cylindrical coordinate system in the ring where $\zeta$ is the coordinate in the direction of propagation around the ring, ranging from 0 to $\mathcal{L}$, so that $\mathcal{L}$ is the ring's circumference. By $\bold{r}_{\perp}$ we denote the pair of coordinates in the plane perpendicular to the direction indicated by $\zeta$. Unlike in the waveguide, here the mode amplitudes $\bold{d}_J(\bold{r_\perp},\zeta)$ can depend on $\zeta$ because if the field's polarization has any component in the plane of the chip, then the Cartesian components of $\bold{d}_J(\bold{r_\perp},\zeta)$ will vary with $\zeta$ \cite{tutorial}.

To model nonlinear interactions in the ring, we expand Eq. \eqref{eq:HNL} with the fields inside the ring; we neglect nonlinear effects in the bus waveguide because the fields' intensity in this region will be much lower than in the ring. We will again introduce nonlinear parameters $\gamma_{NL}$ as defined in \eqref{eq:general_gamma}; the effective areas in the ring are defined in Appendix \ref{appendix:a_eff}. 

In principle, one can solve the Heisenberg equations of motion to derive the evolution of the ring and waveguide fields. However, we instead opt to construct asymptotic field expansions for the fields throughout the system. This enables us to `hide' the full linear dynamics in the field expansions, and proceed with the rest of the TOPDC calculation exactly as we do in Section \ref{section:WG}.

\subsection{Asymptotic Fields}

In this approach, we exploit the fact that a generic displacement field can be expressed using an asymptotic-in or -out expansion \cite{Bethe_asyfields}. These fields are characterized by their asymptotic behaviour. In the ring-channel system with a phantom channel to model loss, an asymptotic-in wavepacket consists of an incoming wavepacket at the bus waveguide's input at $t\rightarrow-\infty$, and outgoing wavepackets in both the bus waveguide and the phantom channel at $t\rightarrow\infty$; thus at negative enough $z$ in the appropriate channel an asymptotic-in wavepacket at $t\rightarrow-\infty$ will have the form of the field that would propagate were the ring not present, as in Eq. (\ref{eq:Dchannel}) \cite{Liscidini_asyfields}. Similarly, an asymptotic-out wavepacket consists of a single outgoing wavepacket at $t\rightarrow\infty$, and fields incoming from the physical and phantom waveguides at $t\rightarrow-\infty$; thus at positive enough $z$ in the appropriate channel an asymptotic-out wavepacket at $t\rightarrow\infty$ will have the form of the field that would propagate in an isolated channel.

Due to this asymptotic behaviour, the asymptotic-in expansion is a natural choice for the incoming pump and seed fields, which are injected in a single waveguide's input; likewise, the asymptotic-out expansion is suitable for the generated modes that we seek at the system's output.

The general form of the fields using an asymptotic-in (-out) expansion is 
\begin{align}
    \bold{D(r},t) &= \sum_{J} \int dk \hspace{1mm} \bold{D}^{asy-in(out)}_J(\bold{r},k) a^{in(out)}_J(k,t)  + H.c.,\label{eq:Dasy}
\end{align}
where $\bold{D}^{asy-in(out)}_J(\bold{r},k)$ is a piecewise function of the position so that the field is defined throughout the entire structure. The operators $a^{in(out)}_J(k,t)$ are associated with a mode of the field defined for the entire structure, and they have the usual bosonic commutation relations.

The full $\bold{D}^{asy-in(out)}_J(\bold{r},k)$ for a ring-channel system like the one considered have been derived earlier \cite{two_strategies}. Since we restrict our description of nonlinear effects to the ring's coordinates, here we only give the field amplitudes in the ring. The asymptotic-in field amplitude in the ring is
\begin{align}
    \nonumber
    \bold{D}^{asy-in}_J(\bold{r},k) &= -\sqrt{\frac{\hbar \omega_{J0}}{4\pi}} {\bold{d}_{J0}(r_\perp)} e^{i\kappa_J\zeta} F_{J-}(k),\\ & \{r_\perp, \zeta\} \in \textup{ring}
    \label{eq:asy.i.ring}
\end{align}
and the asymptotic-out field amplitude is
\begin{align}
    \nonumber
    \bold{D}^{asy-out}_J(\bold{r},k) &= -\sqrt{\frac{\hbar\omega_{J0}}{4\pi}} \bold{d}_{J0}(r_\perp) e^{i{\kappa_J}\zeta} F_{J+}(k) \\ & \{r_\perp, \zeta\} \in \textup{ring}
    \label{eq:asy.o.ring}.
\end{align}
We have introduced
\begin{align}
    F_{J \pm}(k) = \frac{1}{\sqrt{\mathcal{L}}} \bigg(\frac{ \gamma^*_J}{v_J(K_{J} - k) \pm i \bar{\Gamma}_J}\bigg), \label{eq:F}
\end{align}
the Lorentzian field enhancement factor that arises due to the resonant enhancement in the ring. We use $K_J$ to denote the channel wavenumber at the resonant frequency $\omega_J$; $k$ is a component wavenumber in the waveguide band centered at $K_J$. In deriving (\ref{eq:F}), we assume the linewidth $\bar{\Gamma}$ to be narrow enough that to good approximation $\omega_k = \omega_J + v_J(k-K_J)$. Thus we neglect group velocity dispersion across each ring resonance, but take it into account between different resonances by leaving unspecified the dependence of $K_J$ on $\omega_J$ across different resonances.

\subsection{Self- and cross-phase modulation}
\label{section:PM_ring}

The asymptotic-in and -out fields given above include the full linear dynamics of the ring-channel system; here we will show that for a CW pump, it is straightforward to include SPM and XPM in these field expansions. Having taken nonlinear effects in the bus waveguide to be negligible, it suffices to consider the evolution of the ring operators $b_J$ under $ H_\text{ring} + H_{SPM} + H_{XPM}$. 

For each mode $J$, the term in the equation of motion due $H_{\text{ring}}$ is 

\begin{align}
    -\frac{i}{\hbar}[b_J(t),H_{\text{ring}}(t)]
    &= -i \omega_J b_{J}(t). \label{eq:ring_free}
\end{align}

For $H_{SPM}$ and $H_{XPM}$, we expand the general nonlinear Hamiltonian (\ref{eq:HNL}) using (\ref{eq:Dring}). We focus first on SPM on the pump field; collecting the associated terms from the full nonlinear Hamiltonian, we obtain

\begin{align}
    H_{SPM} = -\frac{1}{2} \hbar^2 \omega_P \frac{1}{\mathcal{L}} \gamma_{SPM} v_P^2 b^{\dagger}_P b^{\dagger}_P b_P b_P.
    \label{eq:HSPM_ring}
\end{align}

The nonlinear parameter is defined as in \eqref{eq:general_gamma}, using the effective area given in \eqref{eq:overlap_twodaggers_ring}. The ring operator associated with the pump resonance then evolves under $H_{SPM}$ according to

\begin{align}
    -\frac{i}{\hbar}[b_P(t),H_{SPM}]
    &= i\hbar \omega_P \frac{1}{\mathcal{L}} \gamma_{SPM} v_P^2 b^{\dagger}_{P}(t) b_{P}(t) b_{P}(t),\\
    &= i\hbar \omega_P \frac{1}{\mathcal{L}} \gamma_{SPM} v_P^2 |\beta_J|^2 b_{P}(t). \label{eq:ring_SPM1},\\
    &= i \gamma_{SPM} v_P P'_P b_{P}(t). \label{eq:ring_SPM2}
\end{align}

In (\ref{eq:ring_SPM1}) we take $<b^{\dagger}_P(t) b_P(t)> = |\beta_P(t)|^2$ and we consider the steady state limit where $|\beta_P(t)|^2 = |\beta_P|^2$; in (\ref{eq:ring_SPM2}) we introduce $P'_P = \hbar \omega_P \frac{1}{\mathcal{L}} v_P |\beta_P|^2$, the power of the pump field in the ring. 

Comparing (\ref{eq:ring_free}) and (\ref{eq:ring_SPM2}), one sees that the effect of SPM on a CW pump in the steady state limit is to shift the resonance frequency $\omega_{P}$ to the `hot' resonance frequency $\tilde{\omega}_P = \omega_{P} - \gamma_{SPM} v_P P'_P$. 
When including $H_{SPM}$ in the asymptotic fields derivation, this substitution ultimately appears in the pump's field enhancement factor: 
\begin{align}
    F_{P \pm}(k) = \frac{1}{\sqrt{\mathcal{L}}} \bigg(\frac{ \gamma^*_P}{v_P(\tilde{K}_{P} - k) \pm i \bar{\Gamma}_P}\bigg), \label{eq:F_SPM}
\end{align}
where we introduce $\tilde{K}_P = K_{P} - \gamma_{SPM}P'_P$ as the channel wavenumber associated with the `hot' resonance frequency. {One could equally write Eq. \eqref{eq:F_SPM} in terms of the cold resonance wavenumber $K_P$ and a shifted wavevector $\bar{k} = k+\gamma_{SPM} P'_P$; then the effect of SPM would be written as a shift on the channel wavevector $k$, similar to Eq. \eqref{eq:SPM_shift} for the waveguide. However, here we opt to treat the effect of SPM as a shift in the resonance frequency, since nonlinear effects are confined to the ring, and the bare wavenumber $k$ in the channel is known.}

As the pump power in the ring $P_P'$ builds up, the resonant wavenumber shifts proportionally. However, the steady state value of $P'_P$ is a function of the CW field's detuning from the resonance frequency, with the latter shifting as the field in the ring builds up. Since we will strictly work in the CW limit in these calculations, it is sufficient to determine the steady state value of $P'_P$, with SPM taken into account in the classical equation of motion for $\beta_P(t)$ \cite{stronglydriven_vernon}  . Eq. (\ref{eq:F_SPM}) can then be used in (\ref{eq:asy.i.ring}) and (\ref{eq:asy.o.ring}) to define the asymptotic-in and -out expansions for the CW pump field under $H_L+H_{SPM}+H_{XPM}$, in the steady state limit. 

We apply the same approach to include XPM on the generated modes by the CW pump. The nonlinear Hamiltonian describing this is 
\begin{align}
    H_{XPM} = - 2 \hbar^2 {\omega_{G}} \frac{1}{\mathcal{L}} \gamma_{XPM} v_P v_G b^{\dagger}_P b^{\dagger}_G b_P b_G,
\end{align}
where $G$ denotes a resonance where photons are generated. The nonlinear constant $\gamma_{XPM}$ is defined as in \eqref{eq:gamma_XPM}, now with the effective area defined in \eqref{eq:overlap_twodaggers_ring}. We consider the evolution of the ring operator $b_G$ under $H_\text{ring} + H_{XPM}$, recalling that the SPM term in the Hamiltonian will be negligible for the weak generated fields. The free evolution is given by (\ref{eq:ring_free}), and the XPM term in the steady state gives 
\begin{align}
    -\frac{i}{\hbar}[b_G(t),H_{XPM}(t)]
    &= 2i \gamma_{XPM} v_G P'_P b_{G}(t). \label{eq:ring_XPM}
\end{align}
We then see that the effect of XPM is to shift the resonance frequencies associated with generation modes from $\omega_G$ to the hot resonance frequencies $\tilde{\omega}_G = \omega_G - 2 \gamma_{XPM} v_G P'_P$. The field enhancement factor for the generated modes becomes 
\begin{align}
    F_{G \pm}(k) = \frac{1}{\sqrt{\mathcal{L}}} \bigg(\frac{ \gamma^*_G}{v_G(\tilde{K}_{G} - k) \pm i \bar{\Gamma}_G}\bigg),  \label{eq:F_XPM}
\end{align}
where we identify $\tilde{K}_G = K_{G} - 2 \gamma_{XPM} P'_P$ as the resonant wavenumber for the hot cavity.

\subsection{Resonant triplet generation}
\label{section:TOPDC_ring}

With the modified field enhancements factors (\ref{eq:F_SPM}) and (\ref{eq:F_XPM}), Eqs. (\ref{eq:asy.i.ring}) and (\ref{eq:asy.o.ring}) are the asymptotic field amplitudes arising from $H_0 = H_L + H_{SPM} + H_{XPM}$. We can now approach the TOPDC calculations as we did in section \ref{section:TOPDC_WG}. We split the dynamics up according to $H = H_0 + H_{G_1 G_2 G_3}$, so that the interaction picture Hamiltonian $H_{G_1 G_2 G_3}^{(I)}(t)$ is obtained by using the asymptotic fields to expand the nonlinear Hamiltonian (\ref{eq:HNL}), from which we collect the terms associated with TOPDC. We use the asymptotic-in expansion for the pump and seed fields, and we take these to be classical CW fields so that we have 
\begin{align}
    \nonumber
    \bold{D}_J(\bold{r},t) &= - \sqrt{\frac{\hbar \omega_{J}}{2}} \bold{d}_{J}(\bold{r}_{\perp},\zeta) e^{i\kappa_{J}\zeta} F_{J-}(\tilde{K}_J + \delta \tilde{K}_J)\\ &\times \phi_J e^{-i(\tilde{\omega}_J - \delta \tilde{\omega}_J) t}+ H.c.
    \label{eq:asyin_J}
\end{align}
Here $J=\{P,S\}$ labels the two possible input modes, and $\phi_J$ is the CW field amplitude in the channel as defined in (\ref{eq:NLform}). We quote the field's frequency and wavenumber in terms of the hot resonance values $\tilde{K}_J$ and $ \tilde{\omega}_J$; we use $\delta \tilde{\omega}_J$ and $\delta \tilde{K}_J$ to denote the CW field's detuning from $ \tilde{\omega}_J$ and $\tilde{K}_J$. Due to the narrowness of the resonances, these can be related by $\delta \tilde{\omega}_J = v_J \: \delta \tilde{K}_J$.

We use the asymptotic-out expansion for generated modes, so that
\begin{align}
    \nonumber
    \bold{D}_G(\bold{r},t) &= -\int dk \: \sqrt{\frac{\hbar \omega_{G}}{4\pi}} \bold{d}_{G}(\bold{r}_{\perp},\zeta) e^{i\kappa_G\zeta} F_{G+}(k) a^{out}_G(k)\\ &\times e^{-i\omega_{Gk} t} + H.c.
    \label{eq:asyout_G}
\end{align}
Only asymptotic-out operators will appear in $H_{G_1 G_2 G_3}^{(I)}(t)$, due to our classical treatment of input fields. We will drop the `out' label from the operators, and it should be understood that the operators $a_J(k)$ in the ring-channel system denote asymptotic-out operators. With $H_{G_1 G_2 G_3}^{(I)}(t)$ defined using these fields, we can derive TOPDC generation rates in the microring system to first order, as outlined in Section \ref{section:TOPDC_WG}

\subsubsection{Spontaneous triplet generation}

We again begin by considering degenerate SpTOPDC with a CW pump field, which can in general be detuned from resonance. {In general, there are a number of energy-conserving configurations in which photons can be generated, but we begin by focusing on the degenerate case, deferring the inclusion of these additional processes}. We put \eqref{eq:asyin_J} and \eqref{eq:asyout_G} into \eqref{eq:HNL}, and collect TOPDC terms with $G_1 = G_2 = G_3 \equiv F$. Then we have

\begin{align}
\nonumber
 & H_{FFF}^{(I)}(t)=-\int dk_{1}dk_{2}dk_{3}M_{FFF}(k_{1},k_{2},k_{3})a_{F}^{\dagger}(k_{1})\\
 &\times a_{F}^{\dagger}(k_{2})a_{F}^{\dagger}(k_{3})e^{-i\Omega_{FFF}(k_{1},k_{2},k_{3})t}+H.c.,\label{eq:HFFF_ring}
\end{align}
with 
\begin{align}
 & \Omega_{FFF}(k_{1},k_{2},k_{3})=(\tilde{\omega}_{P}+\delta \tilde{\omega}_{P})-\omega_{Fk_{1}}-\omega_{Fk_{2}}-\omega_{Fk_{3}}
\end{align}
and
\begin{align}
    \nonumber
    \bar{M}_{FFF}&(k_1,k_2,k_3) = \hbar^2 \omega_{F}
    \frac{1}{3\sqrt{2\pi}^3} \phi_P
    \sqrt{v_Pv_F^3} \gamma_{FFF} \mathcal{L}
    \\
    &\times
    F_{P-}(\tilde{K}_P + \delta \tilde{K}_P)
    F^*_{F+}(k_1)
    F^*_{F+}(k_2)
    F^*_{F+}(k_3),
    \label{eq:M.3PG}
\end{align}
where $\gamma_{FFF}$ is defined as in \eqref{eq:gamma_FFF}, with \eqref{eq:general_gamma} and \eqref{eq:overlap_twodaggers_ring}. 
Putting this Hamiltonian into (\ref{eq:Pwork}) we have 
\begin{align}
    {R}_{FFF}/R_P &= (\gamma_{FFF} \mathcal{L})^2 P_{\text{vac}}^2 \: |F_F(\tilde{K}_F)|^6 |F_P(\tilde{K}_P + \delta \tilde{K}_P)|^2 \label{eq:RFFF_ring}.
\end{align}

Unlike in the non-resonant waveguide, the efficiency here scales with the field enhancement factors  associated with each field in the interaction. We also emphasize that the phase matching condition in this case is contained in the definition of $\gamma_{FFF}$, as can be seen in Appendix \ref{appendix:a_eff}. For degenerate SpTOPDC in this microring system, the vacuum power is given by 
\begin{align}
    P_{\text{vac}} = \frac{\hbar\omega_F}{\tau},
\end{align}
where the generation bandwidth is given by 
\begin{align}
    \tau^{-2} = \frac{1}{2}\left( \frac{\bar{\Gamma}_F^4}{((\tilde{\omega}_P + \delta \tilde{\omega}_P) - 3\tilde{\omega}_F)^2 + 9\bar{\Gamma}_F^2} \right). \label{eq:ring_bw_FFF}
\end{align}
Clearly this is maximized at $\tilde{\omega}_P + \delta \tilde{\omega}_P - 3\tilde{\omega}_F = 0$, which corresponds to the three photons being generated on resonance with the hot cavity; the three photons' average frequency in this case would be $\tilde{\omega}_F$, and $\tilde{\omega}_P + \delta \tilde{\omega}_P = 3\tilde{\omega}_F$ would be ensured by energy conservation.

%
%

In this case, 
the expression for the generation bandwidth reduces to $\tau^{-1} = \bar{\Gamma}_F/\sqrt{18}$. This scaling with $\bar{\Gamma}_F$ is expected; $\bar{\Gamma}_F$ is the parameter limiting the range of frequencies over which the generated photons can only be emitted, since they can only be emitted within the resonance at the fundamental frequency.

We have a simple expression for $\tau^{-1}$ in the resonant system since we neglect group velocity dispersion across each resonance; we effectively assume that the resonant linewidth is a tighter constraint than 
the material's dispersion properties. Were this not the case, $\tau^{-2}$ would be defined by an expression like (\ref{eq:WG_bw2}) with Lorentzian envelopes multiplying the integrand. 

Unlike in the waveguide, here there is an additional resonant effect, thus one has a tradeoff between the vacuum power and the overall field enhancement. The vacuum power increases if the linewidth $\bar{\Gamma}_F$ is increased, but the field enhancement factors $|F_F(\tilde{K}_F)|^2$ decrease (recall Eq. \eqref{eq:F_XPM}); since the generation rate has a higher scaling with the field enhancement factors than the vacuum power, a narrow linewidth at $\omega_F$ is required to maximize the generation bandwidth, despite this limiting the vacuum power.  


In the ring-channel system, one can also consider non-degenerate SpTOPDC configurations, where the photon triplet is distributed among two or three ring resonances. Here we consider a scheme where $G_1=G_2\equiv G$, $G_3 \equiv S$, sketched in Fig. \ref{fig:StTOPDC}a. For this particular non-degenerate SpTOPDC configuration, we have
\begin{align}
\nonumber
 & H_{GGS}^{(I)}(t)=-\int dk_{1}dk_{2}dk_{3}M_{GGS}(k_{1},k_{2},k_{3})a_{G}^{\dagger}(k_{1})\\
 &\times {a}_{G}^{\dagger}(k_{2}){a}_{S}^{\dagger}(k_{3})e^{-i\Omega_{GGS}(k_{1},k_{2},k_{3})t}+H.c.,\label{eq:HGGS_ring}
\end{align}
with 
\begin{align}
 & \Omega_{GGS}(k_{1},k_{2},k_{3})=(\tilde{\omega}_{P} + \delta \tilde{\omega}_P) -\omega_{Gk_{1}}-\omega_{Gk_{2}}-\omega_{Sk_{3}}
\end{align}
and
\begin{align}
    \nonumber
    &{M}_{GGS}(k_1,k_2,k_3) = \frac{\hbar^2}{\sqrt{(2\pi)^3}} 
    (\omega_G^2 \omega_S)^{1/3} 
     \phi_P
    \sqrt{v_P v_G^2 v_S}
    \\
    &\times \gamma_{GGS} \mathcal{L}
    F_{P-}(\tilde{K}_P+\delta \tilde{K}_P)
    F^*_{G+}(k_1)
    F^*_{G+}(k_2) F^*_{S+}(k_3),
\end{align}
where 
\begin{align}
    \gamma_{GGS} = \left(\frac{(\omega_G^2 \omega_S \omega_P)^{1/4}}{(\omega_G^2 \omega_S)^{1/3}}\right) \gamma_{GGSP},
\end{align}
with $\gamma_{GGSP}$ defined according to \eqref{eq:general_gamma} and \eqref{eq:overlap_twodaggers_ring}. From this we find the efficiency to be
\begin{align}
    \nonumber {R}_{GGS}/R_P &= (|\gamma_{GGS}| \mathcal{L})^2 P_{\text{vac}}^2 \: |F_G(\tilde{K}_G)|^4 |F_S(\tilde{K}_S)|^2\\
    &\times
    |F_P(\tilde{K}_P+\delta \tilde{K}_P)|^2. \label{eq:RGGS_ring}
\end{align}
Here again, $\delta \tilde{K}_P$ is the CW pump's detuning from the hot cavity's resonance wavenumber. For this process, the characteristic vacuum power is given by 
\begin{align}
    P_{\text{vac}} &=\nonumber \frac{\hbar (\omega_G^2 \omega_S)^{1/3}}{\tau},\\
    \tau^{-2} &=
    \frac{1}{2} \left(\frac{\bar{\Gamma}_G^2 \bar{\Gamma}_S (2\bar{\Gamma}_G+ \bar{\Gamma}_S)}{((\tilde{\omega}_P + \delta \tilde{\omega}_P) - 2\tilde{\omega}_G - \tilde{\omega}_S)^2 +  (2\bar{\Gamma}_G+ \bar{\Gamma}_S)^2} \right).
    \label{eq:ring_bw_GGS_full}
\end{align}

Aside from the dependence on two distinct linewidths, the generation bandwidth implied by \eqref{eq:ring_bw_GGS_full} has the same general form as \eqref{eq:ring_bw_FFF}. If we consider the limit where resonances $G$ and $S$ are similar in the sense that $\omega_G \approx \omega_S$ and $\bar{\Gamma}_G \approx \bar{\Gamma}_S$, the generation bandwidth in \eqref{eq:ring_bw_GGS_full} reduces to 
\begin{align}
    \tau^{-2} = \frac{1}{2}\left( \frac{3\bar{\Gamma}_G^4}{((\tilde{\omega}_P + \delta \tilde{\omega}_P) - 3\tilde{\omega}_G)^2 + 9\bar{\Gamma}_G^2} \right),
\end{align}
consistent with (\ref{eq:ring_bw_FFF}) up to a factor of 3
. If the pump's detuning is set such that $(\tilde{\omega}_T + \delta \tilde{\omega}_T) - 2\tilde{\omega}_G - \tilde{\omega}_S = 0$ so that \eqref{eq:ring_bw_GGS_full} is maximized, we have 
\begin{align}
    \tau^{-2} = \left( \frac{1}{2} \frac{\bar{\Gamma}_G^2 \bar{\Gamma}_S}{2 \bar{\Gamma}_G + \bar{\Gamma}_S} \right).
\end{align}
Here again the generation bandwidth is clearly limited by the resonant linewidths for the generated modes. 

\subsubsection{Stimulated triplet generation} \label{section:StTOPDC_ring}

We now consider seeding the non-degenerate configuration in Fig. \ref{fig:StTOPDC}a with a classical CW field in mode $S$. We treat the seeded mode classically by taking $a^{\dagger}_S(k) \rightarrow \alpha^*_S(k)$ in Eq. (\ref{eq:HGGS_ring}). We assume the pump and seed to be CW fields, again using $\delta \tilde{\omega}_J$ and $\delta \tilde{K}_J$ to denote the input fields' detuning from their respective resonant frequencies and wavenumbers. We then have the interaction Hamiltonian
\begin{align}
\nonumber
 & H_{GG(S)}^{(I)}(t)=-\int dk_{1}dk_{2}M_{GG(S)}(k_{1},k_{2})a_{G}^{\dagger}(k_{1}) {a}_{G}^{\dagger}(k_{2})\\
 &\times e^{-i\Omega_{GG(S)}(k_{1},k_{2})t}+H.c.,\label{eq:HGG(S)_ring}
\end{align}
with 
\begin{align}
 & \Omega_{GG(S)}(k_{1},k_{2})=(\tilde{\omega}_{P} + \delta \tilde{\omega}_P) - (\tilde{\omega}_S + \delta \tilde{\omega}_S) -\omega_{Gk_{1}}-\omega_{Gk_{2}}
\end{align}
and
\begin{align}
    \nonumber
    &M_{GG(S)}(k_1,k_2) = \frac{\hbar^2}{2\pi} \sqrt{\omega_{G}\omega_{S}}
    \phi^*_S \phi_P \gamma_{GG(S)}\mathcal{L} {\sqrt{v_Pv_Sv_G^2}}
    \\
    &\times
    F_{S-}^*(\tilde{K}_S+\delta \tilde{K}_S) F_{P-}(\tilde{K}_P+\delta \tilde{K}_P)
    F^*_{G+}(k_1)F^*_{G+}(k_2),
\end{align}
using the definition for $\gamma_{GG(S)}$ given in Eq. \eqref{eq:gamma_GGS}. The generation efficiency in this case is 
\begin{align} 
    \nonumber{R}_{GG(S)}/R_P &= (|\gamma_{GG(S)}| \mathcal{L})^2 P_S P_{\text{vac}} |F_G(\tilde{K}_G)|^4 \\ &\times |F_S(\tilde{K}_S+\delta \tilde{K}_S)|^2 
    |F_P(\tilde{K}_P + \delta \tilde{K}_P)|^2. \label{eq:RGG(S)_ring}
\end{align}

Here again the generation rate equation is analogous to Eq. (\ref{eq:RS3WG}), with field enhancement factors now appearing in the resonant case. The vacuum power in this case is
\begin{align}
    P_{\text{vac}} &= \nonumber 
    \frac{\hbar \omega_G}{\tau}\\
    \tau^{-1} &= \left( \frac{2\bar{\Gamma}_G^3}{((\tilde{\omega}_P + \delta \tilde{\omega}_P) - (\tilde{\omega}_S + \delta \tilde{\omega}_S) - 2\tilde{\omega}_G)^2 + 4\bar{\Gamma}_G^2} \right).
\end{align}
As in SpTOPDC, the generation bandwidth is constrained by the ring resonance's linewidth at the frequency $\omega_G$. The resonant SpTOPDC and StTOPDC bandwidths have the same scaling with system parameters, but different constant factors, just as we found in the non-resonant case. Comparing (\ref{eq:RGGS_ring}) and (\ref{eq:RGG(S)_ring}) where $\delta \tilde{K}_S = 0$, we have
\begin{align}
    {R}_{GG(S)} = \frac{\gamma_{GG(S)}}{\gamma_{GGS}} \frac{  P_S P_{\text{vac}}^{\text{stim}}}{(P_{\text{vac}}^{\text{spon}})^2} {R}_{GGS},
\end{align}
where we explicitly label the vacuum powers associated with the spontaneous and stimulated processes, since they are distinct. It is convenient to identify an effective vacuum power
\begin{align}
    \bar{P}_{\text{vac}} \equiv \frac{\gamma_{GGS}}{\gamma_{GG(S)}} \frac{(P_{\text{vac}}^{\text{spon}})^2}{ P_{\text{vac}}^{\text{stim}}},
\end{align}
so that the improvement in the efficiency due to the seed field is given by $P_S/\bar{P}_{\text{vac}}$. Here we have 
\begin{align}
    &\bar{P}_{\text{vac}} = \nonumber \hbar \omega_S \left(\frac{((\tilde{\omega}_P + \delta \tilde{\omega}_P) - (\tilde{\omega}_S + \delta \tilde{\omega}_S) - 2\tilde{\omega}_G)^2 + 4\bar{\Gamma}_G^2}{((\tilde{\omega}_P + \delta \tilde{\omega}_P) - \tilde{\omega}_S - 2\tilde{\omega}_G)^2 + (2\bar{\Gamma}_G + \bar{\Gamma}_S)^2} \right)\\
    &\times
    \left(\frac{\bar{\Gamma}_S(2\bar{\Gamma}_G + \bar{\Gamma}_S)}{4\bar{\Gamma}_G} \right). \label{eq:Pvac_eff}
\end{align}
If pump detuning is chosen to satisfy $(\tilde{\omega}_P + \delta \tilde{\omega}_P) - \tilde{\omega}_S - 2\tilde{\omega}_G = 0$, and the seed field is on resonance, then the effective vacuum power reduces to
\begin{align}
    &\bar{P}_{\text{vac}} = \hbar \omega_S \left(\frac{\bar{\Gamma}_G\bar{\Gamma}_S}{ (2\bar{\Gamma}_G + \bar{\Gamma}_S)} \right)
    ,
\end{align}
an easily quantified parameter.




Finally, we address the doubly stimulated configuration in Fig \ref{fig:StTOPDC}b.  The interaction Hamiltonian in the microring system has the form of Eq. (\ref{eq:HDS3}), now with
\begin{align}
    \Omega_{\bar{G}(SS)} &= (\tilde{\omega}_P + \delta \tilde{\omega}_P) - 2(\tilde{\omega}_S + \delta \tilde{\omega}_S) - \omega_{\bar{G}k},\\
    M_{\bar{G}(SS)}(k) &=   \frac{\hbar^2}{\sqrt{2\pi}} \omega_S (\phi_S^*)^2 \phi_P \gamma_{\bar{G}(SS)} \mathcal{L} \sqrt{v_{\bar{G}} v_Pv_S^2}\\
    \nonumber
    &\times (F_{S-}^*(\tilde{K}_S+\delta \tilde{K}_S))^2 F_{P-}(\tilde{K}_P + \delta \tilde{K}_P) F_{\bar{G}+}^*(k),
\end{align}
with the nonlinear parameter is defined as in Eq. \eqref{eq:gamma_G(SS)} with Eq. \eqref{eq:overlap_threedaggers_ring}. The generation rate is
\begin{align}
    {R}_{\bar{G}(SS)}/R_P &= (|\gamma_{\bar{G}(SS)}| \mathcal{L})^2 P_S^2 |F_S(\tilde{K}_S + \delta \tilde{K}_S)|^4 \label{eq:RGSS_ring}
    \\ \nonumber &\times |F_P(\tilde{K}_P + \delta \tilde{K}_P)|^2 |F_{\bar{G}}(\tilde{K}_{\bar{G}} + \delta \tilde{K}_{\bar{G}})|^2, 
\end{align}
where $\tilde{K}_{\bar{G}} + \delta \tilde{K}_{\bar{G}}$ is the wavenumber of the photons generated in mode $\bar{G}$. The detuning $\delta \tilde{K}_{\bar{G}}$ is determined by energy conservation; we have $\delta \tilde{K}_{\bar{G}} = \frac{1}{v_{\bar{G}}} \delta \tilde{\omega}_{\bar{G}}$, and $ \tilde{\omega}_{\bar{G}} + \delta \tilde{\omega}_{\bar{G}} = (\tilde{\omega}_P + \delta \tilde{\omega}_P) - 2(\tilde{\omega}_S + \delta \tilde{\omega}_S)$.

Again we see that the improvement to the efficiency with respect to StTOPDC is given by $P_S/P_{\text{vac}}$; in this case, the comparison is simplified by the fact that this is a classical process {which is not driven by vacuum fluctuations}, thus there is no need to identify an `effective' vacuum power like the one in Eq. \eqref{eq:Pvac_eff}.


\section{Sample Calculation} \label{section: comparison}

We now calculate the different TOPDC generation rates in two particular sample systems. The sample nonresonant system consists of a silica-clad silicon nitride waveguide 1700 nm wide and 800 nm thick; for the resonant system we assume a ring resonator with the same cross-sectional dimensions and the same materials. 

\begin{figure}[h]
    \centering
    \includegraphics[width=0.48\textwidth]{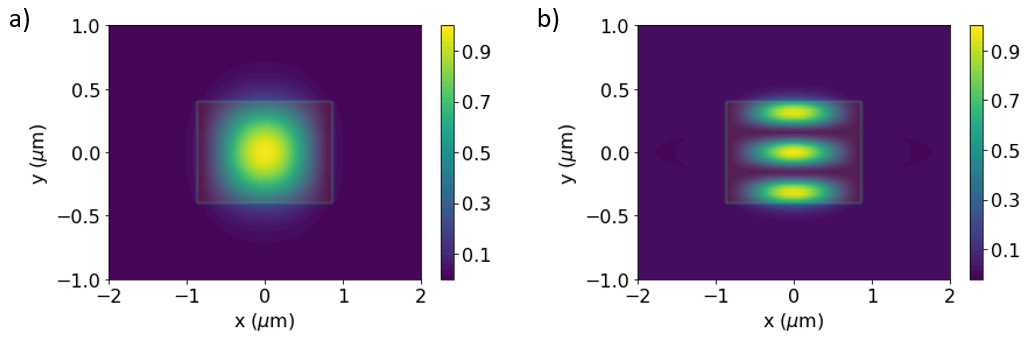}
    \caption{Plots of the electric field intensity $|\bold{E}(x,y)|^2$ in the sample structure for a) the generated modes and b) the pump mode. The rectangle indicates the silicon nitride waveguide; the cladding is silica.}
    \label{fig:Lumericalmodes}
\end{figure}

Phase matching poses a significant challenge in conceiving platforms for TOPDC. In this sample calculation, we use a higher order spatial mode for the pump, keeping the fundamental mode for the generated photons; the pump and generated modes are plotted in Fig. \ref{fig:Lumericalmodes} \cite{Lumerical}.

These plots are generated for $\lambda_P = 0.57$ $\mu$m and $\lambda_{F} = 1.72$ $\mu$m, which are the phase matched wavelengths for degenerate SpTOPDC in the waveguide; the phasematching details for all the processes dicussed here are in Appendix \ref{appendix:calcdetails}. For the other processes, the phase matched wavelengths are slightly different, but they vary so little that the mode profiles do not change significantly.  

With the mode profiles the nonlinear parameter $|\gamma_{G_1,G_2,G_3}|$ can be calculated numerically. For each nonresonant TOPDC process considered, we have $|\gamma_{G_1,G_2,G_3}| = 0.19$ (Wm)$^{-1}$; the nonlinear parameters are all the same because for this waveguide, the mode profiles are essentially unchanged over the small differences in the modes used in each process.

The value of $|\gamma_{G_1,G_2,G_3}|$ for a ring resonator can be different from $|\gamma_{G_1,G_2,G_3}|$ for the corresponding waveguide, depending on the polarization of the light in the ring, and on the radius of curvature of the ring \cite{seifoory_parasitic}. For this sample calculation, however, we take the nonlinear parameter in the waveguide as a good approximation for that in the resonator; we take $|\gamma_{G_1,G_2,G_3}| = 0.19$ (Wm)$^{-1}$ for all the resonant processes. 

\subsection{Degenerate SpTOPDC}

We first consider the waveguide, where from \eqref{eq:RFFF_WG} the TOPDC rate can be written in terms of the incoming pump power $P_P$ as
\begin{align}
    {R}_{FFF}=(|\gamma_{FFF}| L)^{2}P_{\text{vac}}^2\left(\frac{P_P}{\hbar \omega_P}\right), \label{eq:RFFF_WG_2}
\end{align} 
with
\begin{align}
    P_{\text{vac}} &= \frac{\hbar {\omega_F}}{\tau}, \\
    \tau^{-2}_{FFF} &= \frac{v_{F}^{3}}{6\pi^{2}}
 \int dk_{1}dk_{2}dk_{3}\delta(\Omega_{FFF}(k_1,k_2,k_3)) \nonumber\\
 &\times\text{sinc}^{2}\left(\frac{\Delta\bar{k}_{FFF}L}{2}\right),  \label{eq:bw_sample}
\end{align} 
where $\Delta\bar{k}_{FFF}$ is the phase mismatch given in Eq. \eqref{eq:mismatch_Sp}, and $\Omega_{FFF}(k_1,k_2,k_3)$ is the frequency mismatch defined in Eq. \eqref{eq:Omega_FFF}.

For the cross-sectional dimensions described above, we find that degenerate TOPDC is phase matched at $\lambda_P = 0.57 \mu$m and $\lambda_F = 1.72 \mu$m. We take the waveguide's length to be 1 cm, and we calculate the bandwidth by evaluating Eq. \eqref{eq:bw_sample} numerically: We obtain $\Delta \bar{k}$ in the integrand by interpolating simulated dispersion data, and we take the integrals over finite ranges to account for frequency cutoffs in the generated modes (see Appendix \ref{appendix:variablechange}). We find $\tau^{-1} = 2.9\times 10^{4}$ GHz, {which corresponds to $\Delta \lambda \approx 290$ nm.}

The numerical calculation used to obtain this bandwidth can be compared to the analytic expression \eqref{eq:analytictau3PG}
\begin{align}
    \tau^{-2}_{FFF} = \frac{\sqrt{3}}{9|\beta_2|L}, \label{eq:bw_analytic_calc}
\end{align}
which is valid if higher order terms in the dispersion relation can be neglected, and if there is no frequency cutoff for the generated modes. From the dispersion data for the sample waveguide we find $|\beta_2| = 3.2\times10^{-26}$ s$^2$/m, and $\tau^{-1} = 2.4\times10^{4}$ GHz; despite the approximations made in deriving Eq. \eqref{eq:bw_analytic_calc}, the result agrees well with the numerical calculation. {Eq. \eqref{eq:bw_analytic_calc} can be used to easily and accurately estimate the generation bandwidth without the full numerical calculation described in Appendix \ref{appendix:variablechange}}

{In this calculation we have neglected the fact that system losses and detection efficiency may vary over the large generation bandwidth. One way to account for this would be to further restrict the limits of integration in Eq. \eqref{eq:bw_sample} -- for example, to integrate only over frequencies in a particular detection bandwidth. If this were done, Eqs. \eqref{eq:bw_sample} and \eqref{eq:bw_analytic_calc} would no longer agree since the approximations made in deriving Eq. \eqref{eq:bw_analytic_calc} would not apply.} 

{For this sample calculation we simply proceed with $\tau^{-1} = 2.9\times 10^{4}$ GHz, which gives $P_{\text{vac}} = 3.3\times10^{-6}$ W. Using this} and assuming the pump power to be 100 mW, from Eq. \eqref{eq:RFFF_WG_2} we have $R_{FFF} \approx 12$ s$^{-1}$ for the sample waveguide.

For the resonant system, we assume a ring resonator with a 120 $\mu$m radius ($\mathcal{L} \approx 750 \mu$m). For the ring's quality factors we take $Q_F = 10^7$ and $Q_P = 10^5$; the quality factor at the third harmonic frequency is taken to be lower due to the use of a higher order spatial mode. The phase matching condition $\kappa_P - 3 \kappa_F = 0$ is satisfied by the same $\lambda_P$ and $\lambda_F$ used in the waveguide calculation. 

From \eqref{eq:RFFF_ring}, the degenerate SpTOPDC rate in the resonator is given by 
\begin{align}
    {R}_{FFF} &= (\gamma_{FFF} \mathcal{L})^2 P_{\text{vac}}^2 \: |F_F(\tilde{K}_F)|^6 |F_P(\tilde{K}_P + \delta \tilde{K}_P)|^2 \left(\frac{P_P}{\hbar \omega_P}\right), \label{RFFF_sample_1}
\end{align}
where
\begin{align}
    &P_{\text{vac}}^2 = {\hbar^2\omega_F^2} \frac{1}{2}\left( \frac{\bar{\Gamma}_F^4}{((\tilde{\omega}_P + \delta \tilde{\omega}_P) - 3\tilde{\omega}_F)^2 + 9\bar{\Gamma}_F^2} \right),\\
    &|F_T(\tilde{K}_P + \delta \tilde{K}_P)|^2 = \frac{1}{\mathcal{L}} \bigg(\frac{|\gamma_P|^2}{\delta \tilde{\omega}_P^2 + \bar{\Gamma}_P^2}\bigg).
\end{align}
In addition to satisfying phase matching, we seek to maximize the vacuum power and the field enhancement factor for the pump mode . One approach (see Appendix \ref{appendix:calcdetails}) is to set $\delta \tilde{\omega}_P$ such that $\tilde{\omega}_P + \delta \tilde{\omega}_P - 3\tilde{\omega}_F \approx 0$. As discussed in Section \ref{section:ring}, this ensures that the photons are generated on resonance with the hot resonator, maximizing the vacuum power; this would require $\delta \tilde{\omega}_P/2\pi \approx 30$ MHz for the parameters considered here with $P_P = 100$ mW. The pump field is then detuned from resonance, but in this case the linewidth for the pump mode is large enough that the detuning is relatively insignificant, and to good approximation we can put $|F_P(\tilde{K}_P + \delta \tilde{K}_P)|^2 \approx |F_P(\tilde{K}_P)|^2$. Under these conditions, and using $\bar{\Gamma}_J = \omega_J/2 Q_J$, Eq. \eqref{RFFF_sample_1} can be rewritten as
\begin{align}
    {R}_{FFF} &\rightarrow (\gamma_{FFF} \mathcal{L})^2 P_{\text{vac}}^2 \: |F_F(\tilde{K}_F)|^6 |F_P(\tilde{K}_P)|^2 \left(\frac{P_P}{\hbar \omega_P}\right).
\end{align}
where 
\begin{align}
    |F_{J\pm}(\tilde{K}_J)|^2 \rightarrow \frac{1}{\mathcal{L}} \frac{|\gamma_J|^2}{\bar{\Gamma}_J^2}  = \frac{4 v_J \eta_J Q_J}{\mathcal{L} \omega_J},
\end{align}
and
\begin{align}
    P_{\text{vac}}^2 &= \frac{\hbar^2\omega_F^2}{\tau^2},\\
    \tau^{-2} &= \frac{1}{18} \left(\frac{\omega_F}{2 Q_F}\right)^2.
\end{align}

For the parameters outlined above, the generation bandwidth for the resonator is $\tau^{-1} = 1.3\times10^{-2}$ GHz {($\Delta \lambda \approx 0.1$ pm)}, and $P_{\text{vac}} = 1.5 \times 10^{-12}$ W. The generation bandwidth here is much smaller than the bandwidth in the waveguide. This is expected, since the frequency range over which photons can be generated in the resonator is limited by the resonance linewidth; this is a tighter constraint than the material's dispersion properties, which limits the generation bandwidth in the waveguide. Since SpTOPDC scales quadratically with the vacuum power, SpTOPDC in the ring resonator is inefficient despite the enhanced pump power in the ring. Assuming again a 100 mW pump in the channel, which corresponds to 1.05 W in the ring, the rate of triplet generation is $R_{FFF} = 5.9\times10^{-3}$ s$^{-1}$, orders of magnitude smaller than the rate of triplets predicted for the waveguide.
{Clearly a resonant structure would not be ideal for SpTOPDC, even with improvements in parameters; instead, its implementation in a waveguide should be prioritized.}

\subsection{StTOPDC}

As in Sections \ref{section:WG} and \ref{section:ring}, in our discussion of StTOPDC we consider the case where the photon triplet is distributed such that two are emitted in a mode labeled `G', and one is emitted in a seeded mode `S' (see Fig. \ref{fig:StTOPDC}a). For our sample calculation we choose $\lambda_P = 0.57 \mu$m, $\lambda_S = 2.3 \mu$m, and $\lambda_G = 1.52 \mu$m; these wavelengths are phase matched (see Appendix \ref{appendix:calcdetails}), and they result in photon pairs generated in a convenient frequency range for detection. From Eq. \eqref{eq:RS3WG}, the rate of StTOPDC in a waveguide can be written as
\begin{align}
 & {R}_{GG(S)}=(\gamma_{GG(S)}L)^2 P_S P_{\text{vac}} \left( \frac{P_P}{\hbar \omega_P} \right), \label{eq:RS3WG_2}
\end{align}
where
\begin{align}
    P_{\text{vac}} &=  \frac{\hbar \omega_G}{\tau},\\
    \tau^{-1} &= \frac{v_{G}^{2}}{\pi} \int dk_{1}dk_{2}\delta(\Omega_{GG(S)}(k_1,k_2))\:\text{sinc}^{2}\left(\frac{\Delta \bar{k}_{GG(S)} L}{2}\right)
\end{align}
Computing the bandwidth numerically, we find $\tau^{-1} = 4.0\times10^{4}$ GHz {($\Delta \lambda \approx 310$ nm)} and $P_{\text{vac}} = 5.3\times10^{-6}$ W; here again we can compare to the analytic expression
\begin{align}
    \tau_{GG(S)} = \frac{4}{3}\sqrt{\frac{2}{\pi|\beta_2|L}}, \label{eq:analytic_tau_2} 
\end{align}
which is obtained by assuming the modes have no frequency cutoff, and by working up to third order in the dispersion relation (see Section \ref{Section:StTOPDC_WG}). the group velocity dispersion here is slightly different than in SpTOPDC since pairs are generated at a different frequency; we have $|\beta_2| = 5.5\times 10^{-26}$ s$^2$/m, so Eq. \eqref{eq:analytic_tau_2} predicts $\tau^{-1} = 4.5 \times 10^{4}$ GHz, in good agreement with the numerical result. Keeping $P_P = 100$ mW and setting $P_S = 10$ mW, we have $R_{GG(S)} = 5.7\times10^4$ s$^{-1}$. 

For the resonator we consider the analogous scenario to the one described above and in Appendix \ref{appendix:calcdetails}: We maximize the vacuum power by setting $\delta \tilde{\omega}_S=0$, and setting the pump detuning such that $\tilde{\omega}_P + \delta \tilde{\omega}_P - 2\tilde{\omega}_G - \tilde{\omega}_S \approx 0$; since $\tilde{\omega_G} + \tilde{2\omega_S} \approx 3\tilde{\omega}_F$, the required pump detuning with $P_P = 100$ mW is again $\delta \tilde{\omega}_P \approx 30$ MHz, which is a negligible detuning from the pump resonance frequency due to the large linewidth. From Eq. \eqref{eq:RGG(S)_ring}, the rate of StTOPDC in the ring in this limit is given by 
\begin{align} 
    \nonumber
    {R}_{GG(S)} &\rightarrow (|\gamma_{GG(S)}| \mathcal{L})^2 P_S P_{\text{vac}} |F_G(\tilde{K}_G)|^4 \\ &\times |F_S(\tilde{K}_S)|^2 
    |F_P(\tilde{K}_P)|^2\left( \frac{P_P}{\hbar \omega_P} \right).
\end{align}

with 

\begin{align}
    P_{\text{vac}} &= \nonumber 
    \frac{\hbar \omega_G}{\tau}\\
    \tau^{-1} &= \frac{\bar{\Gamma}_G}{2} = \frac{\omega_G}{4 Q_G}.
\end{align}

We assume that $\omega_S$ and $\omega_G$ are sufficiently close to $\omega_F$ that we can take  $Q_S = Q_G = Q_F$. The generation bandwidth then evaluates to $\tau^{-1} = 3.1\times10^{-2}$ GHz {($\Delta \lambda \approx 0.2$ pm)} and $P_{\text{vac}} = 4.0\times10^{-12}$ W. The generation bandwidth in the resonator remains much smaller than in the waveguide. However, the scaling of $R_{GG(S)}$ with the vacuum power is linear, rather than quadratic, so the low vacuum power does not affect the efficiency of the resonator as much. Indeed, the field enhancement in the resonator makes it more efficient than the waveguide for StTOPDC: Setting $P_P = 100$ mW and $P_S = 20 \; \mu$W, we have $R_{GG(S)} = 2.3\times10^5$ s$^{-1}$. We have chosen $P_S = 20 \; \mu$W so that the seed power in the ring is 10 times smaller than the pump power in the ring; recall that for this system $Q_S = 100 Q_P$. 

\subsection{DStTOPDC}

When TOPDC is seeded, alongside StTOPDC there can be a doubly stimulated process where a light is generated in a single mode labeled $\bar{G}$. 
With the pump and seed modes identified above, this mode's wavelength must be $\lambda_{\bar{G}} = 1.12\mu$m to satisfy energy conservation and phase matching. With these conditions satisfied, from Eq. \eqref{eq:RDS3WG} the rate of DStTOPDC in the waveguide is given by 
\begin{align}
 {R}_{\bar{G}(SS)}&= \frac{1}{2\pi}(|\gamma_{\bar{G}(SS)}| L P_{S})^{2} \left( \frac{P_P}{\hbar\omega_P}\right). \label{eq:RDS3WG_2}
\end{align}
With $P_P = 100$ mW and $P_S = 10$ mW, we have $R_{\bar{G}(SS)} = 1.8\times10^{7}$ s$^{-1}$ for the waveguide. 

For the resonator we have
\begin{align}
    {R}_{\bar{G}(SS)}&= (|\gamma_{\bar{G}(SS)}| \mathcal{L})^2 P_S^2 |F_S(\tilde{K}_S + \delta \tilde{K}_S)|^4
    \\ \nonumber &\times |F_P(\tilde{K}_P + \delta \tilde{K}_P)|^2 |F_{\bar{G}}(\tilde{K}_{\bar{G}} + \delta \tilde{K}_{\bar{G}})|^2 \left( \frac{P_P}{\hbar\omega_P}\right),
\end{align}
with 
\begin{align}
    \delta \tilde{K}_{\bar{G}} &= \frac{1}{v_{\bar{G}}} \left( (\tilde{\omega}_P + \delta \tilde{\omega}_P) - 2(\tilde{\omega}_S + \delta \tilde{\omega}_S) - \tilde{\omega}_{\bar{G}} \right).
\end{align}
Keeping the parameters used in the resonant StTOPDC sample calculation (namely $\delta \tilde{\omega}_S = 0$ and $\delta \tilde{\omega}_P/2\pi \approx 30$ MHz, we have $v_{\bar{G}} \delta \tilde{K}_{\bar{G}} << \bar{\Gamma}_{\bar{G}}$; the detuning is sufficiently small that that $|F_{\bar{G}}(\tilde{K}_{\bar{G}} + \delta \tilde{K}_{\bar{G}})|^2 \approx |F_{\bar{G}}(\tilde{K}_{\bar{G}})|^2$ and the DStTOPDC generation rate is to good approximation
\begin{align}
    {R}_{\bar{G}(SS)} &= \nonumber (|\gamma_{\bar{G}(SS)}| \mathcal{L})^2 P_S^2 |F_S(\tilde{K}_S)|^4
    \\&\times |F_P(\tilde{K}_P)|^2 |F_{\bar{G}}(\tilde{K}_{\bar{G}})|^2 \left( \frac{P_P}{\hbar\omega_P}\right). \label{eq:RGSS_ring_2}
\end{align}
Again taking $P_P = 100$ mW and $P_S = 20 \; \mu$W, Eq. \eqref{eq:RGSS_ring_2} predicts $R_{\bar{G}(SS)} = 1.3\times10^{12}$ s$^{-1}$; $R_{\bar{G}(SS)}$ is independent of the generation bandwidth, so DStTOPDC would be much more efficient in a ring than in a waveguide due to the field enhancement.

\subsection{Summary and comments on scaling}

The TOPDC rates for the sample systems are summarized in Table \ref{tab:summary}. 
\begin{table}[h]
    \centering
    \begin{tabular}{|c||c|c|}
        \hline
        & Waveguide & Resonator \\ \hline \hline \rowcolor{LightCyan} $R_{FFF}$ &  $12$ s$^{-1}$ & $5.9\times10^{-3}$ s$^{-1}$ \\ \hline \rowcolor{LightCyan} 
        $R_{GG(S)}$ & $5.7\times10^{4}$ s$^{-1}$ & $2.3\times10^{5}$ s$^{-1}$ \\ \hline 
        \rowcolor{Gray} 
        $R_{\bar{G}(SS)}$ & $1.8\times10^{7}$ s$^{-1}$ & $1.3\times10^{12}$  s$^{-1}$ \\ \hline 
    \end{tabular}
    \caption{Summary of TOPDC rates. {We highlight in blue and gray the quantum (SpTOPDC and StTOPDC) and classical processes (DStTOPDC) respectively.}}
    \label{tab:summary}
\end{table}
{While these results depend on the specific sets of parameters under consideration, our results clearly indicate that non-resonant platforms are preferable for SpTOPDC, whereas resonators perform better in StTOPDC, with an even larger advantage in DStTOPDC. This is because the efficiency of each TOPDC process scales differently with the vacuum power, which depends on the generation bandwidth. In a resonator, there is a trade-off between the field enhancement and the vacuum power: A higher ring quality factor entails a higher field enhancement but a lower vacuum power, since the generation bandwidth is restricted by the linewidth (see the discussions in Section \ref{section:TOPDC_ring}). On the contrary, in a waveguide the generation bandwidth is determined solely by the waveguide's dispersion properties and the phase matching condition (see Section \ref{section:TOPDC_WG}).}

{
Since the SpTOPDC rate scales quadratically with vacuum power, the low vacuum power in the ring affects its performance significantly, despite the field enhancement terms in Eq. \eqref{RFFF_sample_1}. The StTOPDC rate scales only linearly with vacuum power, so the low vacuum power has less of an effect on the overall performance of the resonator; the resonator performs substantially better than the waveguide for the parameters considered here. Finally, because DStTOPDC is a classical process, its rate does not depend on the vacuum power at all, and the ring performs significantly better than the waveguide because the field enhancement comes at no price.
}

We emphasize that the parameters used to obtain {the rates in Table \ref{tab:summary}} were chosen to be realistic, but arguably conservative. For example, we have calculated TOPDC rates for pump and seed powers that are commonly used \cite{PhysRevLett.122.153906,Levy:11}. Obviously, more optimistic rate estimates are obtained with higher input powers; in particular, we have left significant room to increase the seed power in resonant StTOPDC. With higher input powers one would need to include the effect of SPM and XPM due to the seed, which can be done by following the approaches in Sections \ref{section:WG_PM} and \ref{section:PM_ring} for including phase modulation due to the pump. One would especially need to take care in the resonant case, since here $Q_S > Q_P$ so the circulating seed power $P'_S$ can easily exceed the pump power $P'_P$ for the modes used in this sample calculation. Once the new shifts to the wavenumbers and resonance frequencies are accounted for, the TOPDC rates simply scale linearly with each input power.

{We have also been conservative in our choice of the waveguide length, so one could realistically envisage using a longer waveguide.} To illustrate the scaling, consider the waveguide length being increased to $L=10$ cm and $P_P = 500$ mW respectively. We would then have $R_{FFF} = 5.9 \times 10^2$ s$^{-1}$; note that the rate only scales linearly with $L$, since the vacuum power scales with $L^{-1/2}$. We have again used Eq. \eqref{eq:RFFF_WG_2}, in which loss is neglected. For existing low-loss waveguides this is a good approximation \cite{ElDirani_UltralowLoss}, but to realistically model even longer waveguides, loss would need to be considered. This scenario may seem unreasonably optimistic, but it illustrates how, with improvements in certain experimental parameters, TOPDC generation rates in integrated devices could improve to the point of becoming experimentally viable. 

The scaling of the generation rates with the input powers is trivial; however, because the vacuum power and field enhancement factors both depend on parameters such as length and quality factors, the scaling with these parameters is not immediately obvious from the rate expressions as they are written in Sections \ref{section:WG} and \ref{section:ring}. In Table \ref{tab:scaling} we summarize the scaling of the generation rates with some of the system parameters. {Once the scaling of the resonator rates with the quality factors is identified, one can easily understand the results in Table I; the rates of stimulated TOPDC scale more highly with quality factors, so there is more advantage in using a resonant system.}\\
\begin{table}[H]
    \centering
    \begin{tabular}{|c||q|q|g|}
        \hline
        Waveguide parameters & $R_{FFF}$ & $R_{GG(S)}$ & $R_{\bar{G}(SS)}$ \\ \hline 
        $L$ & $\propto L$ & $\propto L^{3/2}$ & $\propto L^2$\\ \hline \hline 
        Ring parameters & $R_{FFF}$ & $R_{GG(S)}$ & $R_{\bar{G}(SS)}$ \\ \hline
        $Q$ & $\propto Q_P Q_F$ & $\propto Q_P Q_G Q_S$ & $\propto Q_P Q_{\bar{G}} Q_S^2$\\ \hline 
        $\mathcal{L}$ & $\propto \mathcal{L}^{-2}$ & $\propto \mathcal{L}^{-2}$ & $\propto \mathcal{L}^{-2}$\\ \hline
    \end{tabular}
    \caption{Summary of phase-matched TOPDC rates' scaling with system parameters. {We highlight in blue and gray the quantum (SpTOPDC and StTOPDC) and classical processes (DStTOPDC) respectively.}}
    \label{tab:scaling}
\end{table}

While the parameters in Table \ref{tab:scaling} impact the TOPDC rates, the most important parameter is arguably the nonlinear parameter $\gamma$. The nonlinear $\gamma$ in this sample system is particularly small; for comparison, $\gamma \sim$ 1 (Wm)$^{-1}$ is typical for spontaneous four-wave mixing in comparable systems \cite{sciadv.aba9186}. The nonlinear parameter here is small due to the use of a higher order spatial mode for the pump, which results in a relatively small effective area. 

The development of platforms in which phase matching can be achieved, while maintaining a higher mode overlap, will be instrumental in making integrated TOPDC viable; there has already been some progress towards this, for example exploring the use of birefringence \cite{Vernay:21}. There has also been progress in the search for high nonlinearity systems \cite{Placke:20, PRXQuantum.2.010337}, another route towards improving the nonlinear parameter. Since the TOPDC rates scale quadratically with $\gamma$, even a moderate increase in the nonlinearity or the mode overlap could make integrated TOPDC viable. 

\section{Conclusions} \label{section:conclusions}

We have discussed the implementation of TOPDC in integrated photonic structures. We derive equations for the rates of spontaneous TOPDC (SpTOPDC) and stimulated TOPDC (StTOPDC) in a waveguide and a microring resonator, explicitly showing the scaling of the rates with system parameters which continue to improve with progress in the fabrication and design of integrated photonic platforms. 

We have verified that a resonant platform is suitable for StTOPDC; on the other hand, SpTOPDC benefits from a platform where the vacuum fluctuations are not limited by a resonant linewidth, so its implementation in non-resonant systems should be prioritized. To illustrate this we present a sample calculation of the TOPDC rates in a silicon nitride waveguide and microring system, assuming conservative parameters compatible with existing technology. We predict observable StTOPDC rates even in this sample resonant system, which is not optimized for TOPDC. We therefore expect that the demonstration of integrated StTOPDC should be possible in the near term. 

Our outlook for integrated SpTOPDC is similarly optimistic, despite the relatively low predicted rates in our sample calculation. There has already been significant progress in the development of platforms with low losses and high nonlinearity \cite{ElDirani_UltralowLoss,Placke:20, PRXQuantum.2.010337}, and the issue of phase matching over large frequency ranges is being addressed \cite{Vernay:21}. Were a TOPDC platform designed taking advantage of these advances, it would be reasonable to expect a TOPDC rate orders of magnitude higher than those predicted in our sample system. Indeed, with the community's focus on designing platforms for TOPDC, integrated SpTOPDC may soon be observable; in the long term, such progress could lead to SpTOPDC becoming a viable source of non-classical light with characteristics that cannot be achieved with presently available sources \cite{PhysRevA.55.2368, tpg1, PhysRevA.103.013704, PRXQuantum.2.030204}. 

\begin{acknowledgments}
 M.B. acknowledges support from the University of Toronto Faculty of Arts \& Science Top Doctoral Fellowship. M.L. acknowledges support by Ministero dell’Istruzione, dell’ Università e della Ricerca (Dipartimenti di Eccellenza Program (2018–2022)). J.E.S. and M.B. acknowledge support from the Natural Sciences and Engineering Research Council of Canada. 
\end{acknowledgments}

\bibliography{apssamp}

\begin{thebibliography}{30}
\bibitem [{Lum()}]{Lumerical}%
  \BibitemOpen
  \href@noop {} {}\bibinfo {note} {Simulations are done using
  Lumerical.}\BibitemShut {Stop}%
\end{thebibliography}

\onecolumngrid
\appendix
\pagebreak

\section{Evaluating waveguide generation bandwidths}
\label{section:BW}

Here we outline the approach towards evaluating the SpTOPDC and StTOPDC generation bandwidths in the waveguide, in the limit where the integrals can be carried out analytically. We first focus on the SpTOPDC bandwidth given in Eq. (\ref{eq:WG_bw}).

We begin by moving to frequency variables, using (\ref{eq:disp1}) and
\begin{align}
 & dk_{n}=\frac{dk_{n}}{d\omega_{Fk_{n}}}d\omega_{Fk_{n}}\rightarrow\frac{1}{v_{F}}d\omega_{Fk_{n}}\rightarrow\frac{1}{v_{F}}d\omega_{n}.
\end{align}
We then have 
\begin{align}
 \tau_{FFF}^{-2} =\frac{1}{6\pi^{2}}\int d\omega_{1}d\omega_{2}d\omega_{3}\delta(\omega_{P}-\omega_{1}-\omega_{2}-\omega_{3})\:\text{sinc}^{2}\left(\Upsilon_{FFF}-\frac{\left(\Delta(\delta \omega_{1})+\Delta(\delta \omega_{2})+\Delta(\delta \omega_{3})\right)L}{2}\right), \label{eq:bw_omega}
\end{align}
where $\Upsilon_{FFF} = (\bar{k}_P - 3\bar{k}_F)\frac{L}{2}$, $\delta \omega_n = \omega_n  - \omega_F$, and the integrals over $\omega_1$, $\omega_2$, $\omega_3$ range from $0$ to $\infty$; the dispersion terms are contained in 
\begin{align}
    \Delta (\delta \omega_n) = \frac{\partial \omega_n}{\partial \bar{k}}(\delta \omega_n) + \frac{1}{2} \frac{\partial^2 \omega_n}{\partial \bar{k}^2} (\delta \omega_n)^2 + ... \label{eq:delta_delta}
\end{align}
The Dirac delta function now restricts the integration to a plane
in ``$\omega$ space,'' where we introduce three mutually orthogonal
unit vectors $\boldsymbol{\hat{1}}$, $\hat{\boldsymbol{2}},$ and
$\boldsymbol{\hat{3}}$ and write 
\begin{align}
 & \boldsymbol{\omega}=\omega_{1}\hat{\boldsymbol{1}}+\omega_{2}\hat{\boldsymbol{2}}+\omega_{3}\hat{\boldsymbol{3}},\label{eq:omega_vector}
\end{align}
where we take $\boldsymbol{\hat{1}}\times\hat{\boldsymbol{2}}=\hat{\boldsymbol{3}}$, etc. The unit vector orthogonal to the plane that the Dirac delta function in (\ref{eq:bw_omega}) specifies is 
\begin{align}
 & \boldsymbol{\hat{z}=}\frac{1}{\sqrt{3}}\left(\hat{\boldsymbol{1}}+\hat{\boldsymbol{2}}+\hat{\boldsymbol{3}}\right).\label{eq:z_vector}
\end{align}
We construct two other unit vectors $\boldsymbol{\hat{x}}$ and $\hat{\boldsymbol{y}}$
mutually orthogonal to each other and to $\hat{\boldsymbol{z}}$,
taking our set to be 
\begin{align}
 & \boldsymbol{\hat{x}}=-\frac{1}{\sqrt{2}}\hat{\boldsymbol{2}}+\frac{1}{\sqrt{2}}\boldsymbol{\hat{3}},\nonumber \\
 & \hat{\boldsymbol{y}}=\frac{2}{\sqrt{6}}\hat{\boldsymbol{1}}-\frac{1}{\sqrt{6}}\hat{\boldsymbol{2}}-\frac{1}{\sqrt{6}}\hat{\boldsymbol{3}},\label{eq:xyz_vectors}\\
 & \hat{\boldsymbol{z}}=\frac{1}{\sqrt{3}}\hat{\boldsymbol{1}}+\frac{1}{\sqrt{3}}\hat{\boldsymbol{2}}+\frac{1}{\sqrt{3}}\hat{\boldsymbol{3}},\nonumber 
\end{align}
such that $\boldsymbol{\hat{x}}\times\hat{\boldsymbol{y}}=\hat{\boldsymbol{z}},$
etc. Then we can write 
\begin{align}
 & \boldsymbol{\omega}=\Omega_{1}\boldsymbol{\hat{x}}+\Omega_{2}\hat{\boldsymbol{y}}+\Omega_{3}\hat{\boldsymbol{z}}.\label{eq:omega_new}
\end{align}
Since $\omega_{1}=\hat{\boldsymbol{1}}\cdot\boldsymbol{\omega}$,
etc., we have 
\begin{align}
 & \omega_{1}=\frac{2}{\sqrt{6}}\Omega_{2}+\frac{1}{\sqrt{3}}\Omega_{3},\label{eq:omegas}\\
 & \omega_{2}=-\frac{1}{\sqrt{2}}\Omega_{1}-\frac{1}{\sqrt{6}}\Omega_{2}+\frac{1}{\sqrt{3}}\Omega_{3},\nonumber \\
 & \omega_{3}=\frac{1}{\sqrt{2}}\Omega_{1}-\frac{1}{\sqrt{6}}\Omega_{2}+\frac{1}{\sqrt{3}}\Omega_{3},\nonumber 
\end{align}
and since the Jacobian of the transformation is unity we have 
\begin{align}
 & d\omega_{1}d\omega_{2}d\omega_{3}=d\Omega_{1}d\Omega_{2}d\Omega_{3}.\label{eq:change}
\end{align}

Using 
\begin{align}
 & \delta(\omega_{P}-\omega_{1}-\omega_{2}-\omega_{3})=\delta(\omega_{P}-\sqrt{3}\Omega_{3})=\frac{1}{\sqrt{3}}\delta(\frac{\omega_{P}}{\sqrt{3}}-\Omega_{3}),\label{eq:delta_work-1}
\end{align}
and noting that when $\Omega_{3}$ is restricted to $\omega_{P}/\sqrt{3}$
we have 
\begin{align}
 & \omega_{1}\rightarrow \frac{2}{\sqrt{6}}\Omega_{2}+\frac{\omega_P}{3},\\
 & \nonumber \omega_{2}\rightarrow -\frac{1}{\sqrt{2}}\Omega_{1}-\frac{1}{\sqrt{6}}\Omega_{2}+\frac{\omega_P}{3},\nonumber \\
 & \nonumber \omega_{3}\rightarrow \frac{1}{\sqrt{2}}\Omega_{1}-\frac{1}{\sqrt{6}}\Omega_{2}+\frac{\omega_P}{3},\nonumber 
\end{align}
and so 
\begin{align}
 & \delta\omega_{1}\rightarrow \frac{2}{\sqrt{6}}\Omega_{2},\\
 & \delta\omega_{2}\rightarrow -\frac{1}{\sqrt{2}}\Omega_{1}-\frac{1}{\sqrt{6}}\Omega_{2},\nonumber \\
 & \delta\omega_{3}\rightarrow \frac{1}{\sqrt{2}}\Omega_{1}-\frac{1}{\sqrt{6}}\Omega_{2}.\nonumber
\end{align}
We can then write
\begin{align}
 & \tau_{FFF}^{-2}=\frac{\sqrt{3}}{18\pi^{2}}\int d\Omega_{1}d\Omega_{2}\: \text{sinc}^{2}\left(\Upsilon_{FFF}-\frac{\left(\Delta(\frac{2}{\sqrt{6}}\Omega_{2})+\Delta(-\frac{1}{\sqrt{2}}\Omega_{1}-\frac{1}{\sqrt{6}}\Omega_{2})+\Delta(\frac{1}{\sqrt{2}}\Omega_{1}-\frac{1}{\sqrt{6}}\Omega_{2})\right)L}{2}\right). \label{eq:tau_newvariables}
\end{align}
{Here $\Omega_1$ and $\Omega_2$ range over values such that $\Omega_1$, $\Omega_2$, $\Omega_3$ lies on the triangle specified by $\omega_1 + \omega_2 + \omega_3 = \omega_P$ with all the $\omega_i > 0$. However, for parameters introduced in the text the sinc function restricts the contributing region of integration to near the center of the triangle specified above, so we can let $\Omega_1$ and $\Omega_2$ range from $-\infty$ to $\infty$.}

Now we introduce new variables $\Omega$ and $\theta$, 
\begin{align}
 & \Omega_{1}=\Omega\sin\theta,\label{eq:polar}\\
 & \Omega_{2}=\Omega\cos\theta,\nonumber 
\end{align}
so that we have 
\begin{align}
 & \tau_{FFF}^{-2}=\frac{\sqrt{3}}{18\pi^{2}}\int_{0}^{\infty}\Omega d\Omega\\
 & \nonumber \times\int_{0}^{2\pi}d\theta\: \text{sinc}^{2}\left(\Upsilon_{FFF}-\frac{\left(\Delta(\frac{2}{\sqrt{6}}\Omega\cos\theta) +\Delta(-\frac{1}{\sqrt{2}}\Omega\sin\theta-\frac{1}{\sqrt{6}}\Omega\cos\theta)+\Delta(\frac{1}{\sqrt{2}}\Omega\sin\theta-\frac{1}{\sqrt{6}}\Omega\cos\theta)\right)L}{2}\right).
\end{align}
Working up to fourth order in Eq. (\ref{eq:delta_delta}), and noting $\delta \omega_1 + \delta \omega_2 + \delta \omega_3 = 0$, we find
\begin{align}
 & \Delta(\delta \omega_{1})+\Delta(\delta \omega_{2})+\Delta(\delta \omega_{3})\\
 & \nonumber =\frac{1}{2}\beta_{2}(\delta \omega_{1}^{2}+\delta \omega_{2}^{2}+\delta \omega_{3}^{2})+\frac{1}{6}\beta_{3}(\delta \omega_{1}^{3}+\delta \omega_{2}^{3}+\delta \omega_{3}^{3})+\frac{1}{24}\beta_{4}(\delta \omega_{1}^{4}+\delta \omega_{2}^{4}+\delta \omega_{3}^{4})\\
 & \nonumber =\frac{1}{2}\beta_{2}\Omega^{2}+\frac{1}{6\sqrt{6}}\beta_{3}\Omega^{3}\cos3\theta+\frac{1}{48}\beta_{4}\Omega^{4},
\end{align}
and so 
\begin{align}
 & \tau_{FFF}^{-2}=\frac{\sqrt{3}}{18\pi^{2}}\int_{0}^{\infty}\Omega d\Omega\int_{0}^{2\pi}d\theta\:\text{sinc}^{2}\left(\Upsilon_{FFF}-\frac{\left(\frac{1}{2}\beta_{2}\Omega^{2}+\frac{1}{6\sqrt{6}}\beta_{3}\Omega^{3}\cos3\theta+\frac{1}{48}\beta_{4}\Omega^{4}\right)L}{2}\right).
\end{align}

We can introduce a new variable $\mu=3\theta$ which will range from
$0$ to $6\pi$ (i.e., three periods) as $\theta$ ranges from $0$
to $2\pi$; however, $d\theta=d\mu/3$, so for the integrand written
in terms of $\mu$ we can integrate $\mu$ from $0$ to $2\pi$. Further,
we put introduce a new variable $y=\Omega^{2};$ then $\Omega d\Omega=dy/2$,
and we have
\begin{align}
 & \tau_{FFF}^{-2}=\frac{\sqrt{3}}{36\pi^{2}}\int_{0}^{\infty}dy\int_{0}^{2\pi}d\mu\:\text{sinc}^{2}\left(\Upsilon_{FFF}-\frac{\left(\frac{1}{2}\beta_{2}y+\frac{1}{6\sqrt{6}}\beta_{3}y^{3/2}\cos\mu+\frac{1}{48}\beta_{4}y^{2}\right)L}{2}\right).\label{eq:tau3PG}
\end{align}
{In the special case in which we assume phase matching, and negligble higher order dispersion terms, this can be evaluated analytically}; we take $\Upsilon_{FFF} = 0, \beta_3 = 0, \beta_4 = 0$. The dependence in the integrand of (\ref{eq:tau3PG}) on $\mu$ vanishes, and we have
\begin{align}
 & \tau_{FFF}^{-2}=\frac{\sqrt{3}}{18\pi}\int_{0}^{\infty}dy\left(\frac{\sin^{2}\left(\frac{\beta_{2}yL}{4}\right)}{\left(\frac{\beta_{2}yL}{4}\right)^{2}}\right)\label{eq:tauwork1-1}\\
 & =\frac{\sqrt{3}}{9}\frac{1}{\left|\beta_{2}\right|L}.\nonumber 
\end{align}

We process $\tau^{-1}$ for StTOPDC in a similar way. Writing Eq. (\ref{eq:WG_bw}) in terms of frequency, we find
\begin{align}
 & \tau^{-1}_{GG(S)}=\frac{1}{\pi}\int d\omega_{1}d\omega_{2}\delta(\omega_{P}-\omega_{S}-\omega_{1}-\omega_{2})\:\text{sinc}^{2}\left(\Upsilon_{GG(S)}-\frac{\left(\Delta(\delta \omega_{1})+\Delta(\delta \omega_{2})\right)L}{2}\right),\label{eq:RSFWM-1}
\end{align}
where $\Upsilon_{GG(S)}=(\bar{k}_{P}-\bar{k}_{S}-2\bar{k}_{G}) \frac{L}{2}$. We again introduce two new variables, $\omega_{av}$ and $\Omega$, such that 
\begin{align}
 & \nonumber \omega_{1}=\omega_{av}+\frac{1}{2}\Omega,\\
 & \omega_{2}=\omega_{av}-\frac{1}{2}\Omega. \label{eq:St_bw_variables}
\end{align}

The Jacobian is unity, and the Dirac delta in (\ref{eq:RSFWM-1})
here reduces to 
\begin{align}
 & \delta(\omega_{T}-\omega_{S}-2\omega_{av})=\delta(2\omega_{G}-2\omega_{av})=\frac{1}{2}\delta(\omega_{av}-\omega_{G}),
\end{align}
and when $\omega_{av}=\omega_{G}$ we have 
\begin{align}
 & \delta \omega_{1}\rightarrow\frac{1}{2}\Omega,\\
 & \nonumber \delta \omega_{2}\rightarrow-\frac{1}{2}\Omega,
\end{align},
and we have 
\begin{align}
 & \tau_{GG(S)}^{-1}=\frac{1}{2\pi}\int_{-\infty}^{\infty}d\Omega\:\text{sinc}^{2}\left(\Upsilon_{GG(S)}-\frac{\left(\Delta(\frac{1}{2}\Omega)+\Delta(-\frac{1}{2}\Omega)\right)L}{2}\right). \label{eq:tau_St_newvars}
\end{align}

Using the same strategy as above, this can be reduced to
\begin{align}
 & \tau_{GG(S)}^{-1}=\frac{1}{2\pi}\int_{-\infty}^{\infty}d\Omega\:\text{sinc}^{2}\left(\Upsilon_{GG(S)}--\frac{1}{8}\beta_{2}L\Omega^{2}-\frac{1}{384}\beta_{4}L\Omega^{4}\right).
\end{align}
In the limit where $\Upsilon_{GG(S)}=0$
and $\beta_{4}=0$ we have 
\begin{align}
 & \tau_{GG(S)}^{-1}=\frac{4}{3}\sqrt{\frac{2}{\pi\left|\beta_{2}\right|L}}.
\end{align}

\newpage 
\label{appendix:evaluatingtau}

\section{Evaluating bandwidths for a finite frequency range}
\label{appendix:variablechange}

In Appendix \ref{appendix:evaluatingtau} we considered an analytic limit, where one works to the first few orders in the dispersion expansion, and the integrals over $\Omega_1$ and $\Omega_2$ are taken to go from $-\infty$ to $\infty$. In reality, the integration range for the bandwidth will be restricted, either because of frequency cutoffs for the modes associated with the frequency variables, or to account for a finite detection bandwidth. Here we find the ranges of integration for $\Omega_1$ and $\Omega_2$ in Eq. \eqref{eq:tau_newvariables}.

We use $\omega_{\text{min}}$ and $\omega_{\text{\text{max}}}$ to denote the integration limits in the original coordinate system, so  
\begin{align}
    \omega_{\text{min}} \leq \omega_n \leq \omega_{\text{max}}. \label{eq:bw_min_max}
\end{align}
Then with \eqref{eq:omegas} we have
\begin{align}
 & \omega_{\text{min}} \leq \frac{2}{\sqrt{6}}\Omega_{2}+\frac{1}{\sqrt{3}}\Omega_{3} \leq \omega_{\text{max}}\\
 & \omega_{\text{min}} \leq -\frac{1}{\sqrt{2}}\Omega_{1}-\frac{1}{\sqrt{6}}\Omega_{2}+\frac{1}{\sqrt{3}}\Omega_{3} \leq \omega_{\text{max}},\nonumber \\
 & \omega_{\text{min}} \leq \frac{1}{\sqrt{2}}\Omega_{1}-\frac{1}{\sqrt{6}}\Omega_{2}+\frac{1}{\sqrt{3}}\Omega_{3} \leq \omega_{\text{max}},\nonumber 
\end{align}
and setting $\Omega_3 = \omega_P/\sqrt{3}$ (which is imposed by the delta function), we have
\begin{align}
 & \omega_{\text{min}} \leq \frac{2}{\sqrt{6}}\Omega_{2}+\frac{\omega_P}{3} \leq \omega_{\text{max}} \label{eq:Omega_limits}\\
 & \omega_{\text{min}} \leq -\frac{1}{\sqrt{2}}\Omega_{1}-\frac{1}{\sqrt{6}}\Omega_{2}+\frac{\omega_P}{3} \leq \omega_{\text{max}},\nonumber \\
 & \omega_{\text{min}} \leq \frac{1}{\sqrt{2}}\Omega_{1}-\frac{1}{\sqrt{6}}\Omega_{2}+\frac{\omega_P}{3} \leq \omega_{\text{max}}.\nonumber 
\end{align}

We can immediately see that $\Omega_2$ is constrained according to 
\begin{align}
\frac{\sqrt{6}}{2}\omega_{\text{min}} - \frac{\omega_P}{\sqrt{6}} \leq \Omega_{2} \leq \frac{\sqrt{6}}{2}\omega_{\text{max}} - \frac{\omega_P}{\sqrt{6}}
\end{align}
From the latter two expressions in Eq. \eqref{eq:Omega_limits}, one can see that the limits on $\Omega_1$ depend on the particular value of $\Omega_2$. Introducing $\bar{\Omega}_2$ to explicitly denote a specific value of $\Omega_2$, we have two sets of inequalities defining $\Omega_1$:
\begin{align}
  -\sqrt{2}\left(\omega_{\text{min}} - \frac{\omega_P}{3} + \frac{1}{\sqrt{6}}\bar{\Omega}_{2} \right) \geq\: &\Omega_{1} \geq -\sqrt{2}\left(\omega_{\text{max}} - \frac{\omega_P}{3} + \frac{1}{\sqrt{6}}\bar{\Omega}_{2}\right), \\
  \sqrt{2} \left(\omega_{\text{min}} - \frac{\omega_P}{3} + \frac{1}{\sqrt{6}}\bar{\Omega}_{2}\right)\leq \: &\Omega_{1} \leq \sqrt{2} \left( \omega_{\text{max}} - \frac{\omega_P}{3} + \frac{1}{\sqrt{6}}\bar{\Omega}_{2} \right).\nonumber 
\end{align}
The range of $\Omega_1$ will be constrained by the tighter bounds; we can write 
\begin{align}
    \Omega_1 \leq \text{min} \left\{ \sqrt{2} \left( \omega_{\text{max}} - \frac{\omega_P}{3} + \frac{1}{\sqrt{6}}\bar{\Omega}_{2} \right), -\sqrt{2}\left(\omega_{\text{min}} - \frac{\omega_P}{3} + \frac{1}{\sqrt{6}}\bar{\Omega}_{2} \right) \right\}\\
    \Omega_1 \geq \text{max} \left\{ -\sqrt{2} \left( \omega_{\text{max}} - \frac{\omega_P}{3} + \frac{1}{\sqrt{6}}\bar{\Omega}_{2} \right), \sqrt{2}\left(\omega_{\text{min}} - \frac{\omega_P}{3} + \frac{1}{\sqrt{6}}\bar{\Omega}_{2} \right) \right\}.
\end{align}
In summary, the generation bandwidth for a finite frequency range can be written as 
\begin{align}
 & \tau_{FFF}^{-2}=\frac{\sqrt{3}}{18\pi^{2}}\int^{\Omega_{1,\text{max}}(\Omega_2)}_{\Omega_{1,\text{min}}(\Omega_2)} d\Omega_{1} \int^{\Omega_{2,\text{max}}}_{\Omega_{2,\text{min}}} d\Omega_{2}\: \text{sinc}^{2}\left(\Upsilon_{FFF}-\frac{\left(\Delta(\frac{2}{\sqrt{6}}\Omega_{2})+\Delta(-\frac{1}{\sqrt{2}}\Omega_{1}-\frac{1}{\sqrt{6}}\Omega_{2})+\Delta(\frac{1}{\sqrt{2}}\Omega_{1}-\frac{1}{\sqrt{6}}\Omega_{2})\right)L}{2}\right). \label{eq:tau_newbounds}
\end{align}
where 
\begin{align}
    & \Omega_{2,\text{min}} = \frac{\sqrt{6}}{2}\omega_{\text{min}} - \frac{\omega_P}{\sqrt{6}} \\
    & \Omega_{2,\text{max}} = \frac{\sqrt{6}}{2}\omega_{\text{max}} -  \frac{\omega_P}{\sqrt{6}}\\
    & \Omega_{1,\text{min}}(\Omega_2) = \text{max} \left\{ -\sqrt{2} \left( \omega_{\text{max}} - \frac{\omega_P}{3} + \frac{1}{\sqrt{6}}\Omega_{2} \right), \sqrt{2}\left(\omega_{\text{min}} - \frac{\omega_P}{3} + \frac{1}{\sqrt{6}}\Omega_{2} \right) \right\} \\
    & \Omega_{1,\text{max}}(\Omega_2) = \text{min} \left\{ \sqrt{2} \left( \omega_{\text{max}} - \frac{\omega_P}{3} + \frac{1}{\sqrt{6}}\Omega_{2} \right), -\sqrt{2}\left(\omega_{\text{min}} - \frac{\omega_P}{3} + \frac{1}{\sqrt{6}}\Omega_{2} \right) \right\},
\end{align}
which can be implemented numerically, as we do in the sample calculation in Section \ref{section: comparison}.

We approach the StTOPDC calculation similarly. The bandwidth is given in terms of frequency by Eq. \eqref{eq:RSFWM-1}, and we adopt the transformed variables defined in Eq. \eqref{eq:St_bw_variables}. From Eq. \eqref{eq:bw_min_max}, the values of the new variables $\Omega$ and $\omega_{av}$ are constrained by
\begin{align}
     & \omega_{\text{min}} \leq \omega_{av} + \frac{1}{2} \Omega \leq \omega_{\text{max}}\\
     & \omega_{\text{min}} \leq \omega_{av} - \frac{1}{2} \Omega \leq \omega_{\text{max}},\nonumber
\end{align}
from which we obtain
\begin{align}
    \text{max}\left\{ \omega_{\text{min}} - \frac{1}{2} \Omega, \omega_{\text{min}} + \frac{1}{2} \Omega \right\} \leq \omega_{av} \leq \text{min} \left\{ \omega_{\text{max}} - \frac{1}{2} \Omega, \omega_{\text{max}} + \frac{1}{2} \Omega \right\}.
\end{align}
It is sufficient to consider the largest possible range for $\omega_{av}$, which is
\begin{align}
    \omega_{\text{min}} \leq \omega_{\text{max}},
\end{align}
which occurs when $\Omega = 0$. 

The range of $\Omega$ can be written in terms of a particular value of $\omega_{av}$ as 
\begin{align}
    \text{max}\left\{ 2(\omega_{\text{min}} - \omega_{av}), -2(\omega_{\text{max}} - \omega_{av}) \right\} \leq \Omega \leq \text{min} \left\{ 2(\omega_{\text{max}} - \omega_{av}), -2(\omega_{\text{min}} - \omega_{av}) \right\}.
\end{align}
As discussed in Appendix \ref{appendix:evaluatingtau}, the delta function in \eqref{eq:RSFWM-1} sets $\omega_{av}=\omega_{G}$, and with the finite integration limits we have a slight modification to \eqref{eq:tau_St_newvars}; we have
\begin{align}
 & \tau_{GG(S)}^{-1}=\frac{1}{2\pi}\int_{\Omega_{\text{min}}}^{\Omega_{\text{max}}}d\Omega\:\text{sinc}^{2}\left(\Upsilon_{GG(S)}-\frac{\left(\Delta(\frac{1}{2}\Omega)+\Delta(-\frac{1}{2}\Omega)\right)L}{2}\right),
\end{align}
where 
\begin{align}
 \Omega_{\text{max}} &= \text{min} \left\{ 2(\omega_{\text{max}} - \omega_{G}), -2(\omega_{\text{min}} - \omega_{G}) \right\} \\
 \Omega_{\text{min}} &= \text{max}\left\{ 2(\omega_{\text{min}} - \omega_{G}), -2(\omega_{\text{max}} - \omega_{G}) \right\},
\end{align}
which can be computed numerically for a particular set of frequencies and dispersion data.

\newpage

\section{Sample calculation details}

\label{appendix:calcdetails}

\begin{figure}[h]
    \centering
    \includegraphics[width=0.5\textwidth]{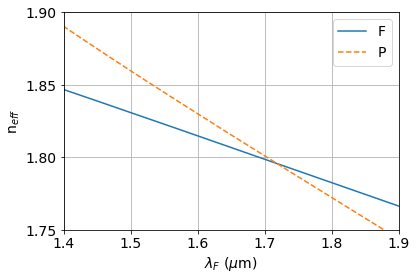}
    \caption{Effective indices for the fundamental (solid blue line) and third harmonic (dashed orange line) modes used in the denegerate SpTOPDC sample calculation. We have used $\lambda_{P} = \lambda_F/3$ to plot both the fundamental and third harmonic indices against a single axis.}
    \label{fig:Sp_phasematch}
\end{figure}

Material data and waveguide dispersion is obtained by simulation in Lumerical using the default material properties; for silicon nitride we use the dataset from Phillip \cite{Lumerical}. In Fig. \ref{fig:Sp_phasematch} we show the simulated dispersion plots for the fundamental and third harmonic modes in the sample structure; phase matching is achieved at $\lambda_F \approx 1.72 \mu$m, with $\lambda_{P} = \lambda_F/3 \approx 0.57 \mu$m.

When seeking phase matched wavelengths for nondegenerate TOPDC processes, we simply choose a desired separation of the generation modes and compute the phase mismatch for different pump frequencies; we denote this separation with $\Delta$ in Fig. \ref{fig:StTOPDC}. We find that with $\Delta$ chosen such that $\lambda_G \approx 1.52 \mu$m and $\lambda_S \approx 2.3 \mu$m, the change in the pump frequency is negligible and we can keep $\lambda_P = 0.57\mu$m while satisfying phase matching. For the DStTOPDC calculation, we keep these values $\lambda_P$ and $\lambda_S$, and seek the value of $\lambda_{\bar{G}}$ which minimizes the phase mismatch; DStTOPDC is phase matched with $\lambda_{\bar{G}}\approx 1.12 \mu$m. 

\subsection{Self- and cross-phase modulation}

Here we verify that self- and cross-phase modulation do not have a significant effect on phase matching. Using the effective index data shown in Fig. \ref{fig:Sp_phasematch} we find the "bare" wavevectors $k_P = 3k_F = 1.97 \times 10^7$ m$^{-1}$. With cross-phase modulation, for phase matching we require $\bar{k}_P = 3\bar{k}_F$ where 

\begin{align}
    \bar{k}_P &= k_P + \gamma_{SPM} P_P\\
    \bar{k}_F &= k_F + 2\gamma_{XPM} P_P.
\end{align}

Using the simulated mode profiles \cite{Lumerical}, we calculate $\gamma_{SPM} \approx 4.3$ (Wm)$^{-1}$ and $\gamma_{XPM} \approx 0.8$ (Wm)$^{-1}$. For the maximum pump power considered in this manuscript, $P_P = 100$ mW, the shift to the wavevectors is at least seven orders of magnitude smaller than the wavevectors themselves. Even if the pump power were increased to 10 W, the shift to the wavevectors would be small enough that the shift in $\lambda_P$ and $\lambda_F$ required to maintain phase matching is be negligible; the phase matched wavelengths identified above are valid with SPM and XPM accounted for. 

A similar argument applies to the non-degenerate configurations; the magnitude of the shifts due to SPM and XPM remains much smaller than the bare wavevectors, so the effect of SPM and XPM on the phase matching is negligible.

\subsection{Resonant TOPDC}

In the resonant system, the generation rate is affected not only by the phase mismatch but also by the detuning from resonance. Here we discuss the maximization of the resonant TOPDC rates. 

\subsubsection{Spontaneous TOPDC}

The TOPDC rate is given by (see Eq. \eqref{eq:RFFF_ring})
\begin{align}
    {R}_{FFF}/R_P &= (\gamma_{FFF} \mathcal{L})^2 P_{\text{vac}}^2 \: |F_F(\tilde{K}_F)|^6 |F_P(\tilde{K}_P + \delta \tilde{K}_P)|^2,
\end{align}
where
\begin{align}
    &P_{\text{vac}}^2 = {\hbar^2\omega_F^2} \frac{1}{2}\left( \frac{\bar{\Gamma}_F^4}{((\tilde{\omega}_P + \delta \tilde{\omega}_P) - 3\tilde{\omega}_F)^2 + 9\bar{\Gamma}_F^2} \right),\\
    &|F_T(\tilde{K}_P + \delta \tilde{K}_P)|^2 = \frac{1}{\mathcal{L}} \bigg(\frac{|\gamma_P|^2}{\delta \tilde{\omega}_P^2 + \bar{\Gamma}_P^2}\bigg).
\end{align}
We focus first on maximizing the vacuum power. This requires setting
\begin{align}
    \delta \tilde{\omega}_P &= 3\tilde{\omega}_F - \tilde{\omega}_P\\ 
    &= 3 \omega_F - \omega_P - (6\gamma_{XPM} v_F - \gamma_{SPM} v_P) P'_P\\
    &= - (6\gamma_{XPM} v_F - \gamma_{SPM} v_P) P'_P,
\end{align}
where we have verified that $\omega_P = 3\omega_F$ for the phase matched modes. We have $\gamma_{SPM} = 4.3$ (Wm)$^{-1}$ and $\gamma_{XPM} = 0.8$ (Wm)$^{-1}$, and the group indices are $n_{g(F)} \approx 2.1$, $n_{g(P)} \approx 2.3$ from which we obtain the group velocities \cite{Lumerical}. With $P_P = 100$ mW the field enhancement yields $P'_P = 1.05$ W, and we require $\delta \tilde{\omega}_P/2\pi \approx$ 30 MHz. 

Because of the large linewidth of the pump resonance, the effect of the detuning on the field enhancement factor is negligible; we take $Q_P = 10^5$ which implies $\bar{\Gamma}_P/2\pi = 2.6$ GHz. Since the linewidth is orders of magnitude higher than the detuning, even with the detuning we have $|F_P(\tilde{K}_P + \delta \tilde{K}_P)|^2 \approx |F_P(\tilde{K}_P)|^2$, and to good approximation the TOPDC rate can be written as 
\begin{align}
    {R}_{FFF}/R_P &= (\gamma_{FFF} \mathcal{L})^2 P_{\text{vac}}^2 \: |F_F(\tilde{K}_F)|^6 |F_P(\tilde{K}_P)|^2,\\
    P_{\text{vac}}^2 &=\hbar^2\omega_F^2 \frac{\bar{\Gamma}_F^2}{18}.
\end{align}

\subsubsection{Stimulated TOPDC}

For resonant StTOPDC we have (see Eq. \eqref{eq:RGG(S)_ring})
\begin{align}
    {R}_{GG(S)}/R_P &= (|\gamma_{GG(S)}| \mathcal{L})^2 P_S P_{\text{vac}} |F_G(\tilde{K}_G)|^4 |F_S(\tilde{K}_S+\delta \tilde{K}_S)|^2 
    |F_P(\tilde{K}_P + \delta \tilde{K}_P)|^2,\\
     P_{\text{vac}} &= {\hbar \omega_G}\left( \frac{2\bar{\Gamma}_G^3}{((\tilde{\omega}_P + \delta \tilde{\omega}_P) - (\tilde{\omega}_S + \delta \tilde{\omega}_S) - 2\tilde{\omega}_G)^2 + 4\bar{\Gamma}_G^2} \right).
\end{align}
Here we set $\delta \tilde{\omega}_S=0$. For the StTOPDC modes we have $\tilde{\omega}_S + 2\tilde{\omega}_G \approx 3\tilde{\omega}_F$, so here too we can maximize the vacuum power by setting $\delta \tilde{\omega}_P/2\pi \approx 30$ MHz. As discussed above, this detuning has a negligible effect on the pump field enhancement due to the large linewidth of the pump resonance. The same approach applies for non-degenerate SpTOPDC (see \eqref{eq:RGGS_ring}).

For DStTOPDC there is no vacuum power to maximize; we have (see Eq. \eqref{eq:RGSS_ring})
\begin{align}
    {R}_{\bar{G}(SS)}/R_P &= (|\gamma_{\bar{G}(SS)}| \mathcal{L})^2 P_S^2 |F_S(\tilde{K}_S + \delta \tilde{K}_S)|^4 |F_P(\tilde{K}_P + \delta \tilde{K}_P)|^2 |F_{\bar{G}}(\tilde{K}_{\bar{G}} + \delta \tilde{K}_{\bar{G}})|^2, \\
    \delta \tilde{K}_{\bar{G}} &= \frac{1}{v_{\bar{G}}} \left( (\tilde{\omega}_P + \delta \tilde{\omega}_P) - 2(\tilde{\omega}_S + \delta \tilde{\omega}_S) - \tilde{\omega}_{\bar{G}} \right).
\end{align}
If the goal is to maximize ${R}_{\bar{G}(SS)}$, here again we can we set $\delta \tilde{\omega}_S=0$ and put $\delta \tilde{\omega}_P/2\pi \approx 30$ MHz to obtain $\delta \tilde{K}_{\bar{G}} = 0$. 

\newpage

\section{Effective areas}
\label{appendix:a_eff}

\subsection{Waveguide}

For nonlinear processes involving two raising operators and two lowering operators, we have
\begin{align}
    \frac{e^{i\Phi_{J1,J2,J3,J4}}}{\mathcal{A}_{J1,J2,J3,J4}} = \frac{\int dxdy (\chi_{3}^{ijkl}(x,y)/\bar{\chi}_{3}) e^{*i}_{J1}(x,y)e^{*j}_{J2}(x,y)e^k_{J3}(x,y)e^l_{J4}(x,y)}{\mathcal{N}_{J1}\mathcal{N}_{J2}\mathcal{N}_{J3}\mathcal{N}_{J4}},
    \label{eq:overlap_twodaggers}
\end{align}
where
\begin{align}
    e^i_J(x,y) = \frac{1}{\epsilon_0 \varepsilon_1(x,y;\omega_J)} d^i_J(x,y)
\end{align}
is a component of the electric field mode profile, and the $\mathcal{N}_{J}$ are normalization constants defined as 
\begin{align}
    \mathcal{N}_J = \sqrt{\int dxdy \: \bold{e}_J(x,y)\cdot \bold{e}_J(x,y)\frac{n(x,y;\omega_J)/\bar{n}_J}{v_g(x,y;\omega_J)/v_J}}.
\end{align}

We use a similar definition for the effective area in processes with three raising operators; in this case we have 
\begin{align}
    \frac{e^{i\Phi_{J1,J2,J3,J4}}}{\mathcal{A}_{J1,J2,J3,J4}} = \frac{\int dxdy (\chi_{3}^{ijkl}/\bar{\chi}_{3}) e^{*i}_{J1}(x,y)e^{*j}_{J2}(x,y)e^{*k}_{J3}(x,y)e^l_{J4}(x,y)}{\mathcal{N}_{J1}\mathcal{N}_{J2}\mathcal{N}_{J3}\mathcal{N}_{J4}}.
    \label{eq:overlap_threedaggers}
\end{align}

\subsection{Ring resonator}

In the ring, the effective area for a process involving two raising operators is
\begin{align}
    \frac{e^{i\Phi_{J1,J2,J3,J4}}}{\mathcal{A}_{J1,J2,J3,J4}} =\frac{1}{\mathcal{L}} \frac{\int d\bold{r}_{\perp}d\zeta (\chi_{3}^{ijkl}(\bold{r}_{\perp})/\bar{\chi}_{3}) e^{*i}_{J1}(\bold{r_\perp},\zeta)e^{*j}_{J2}(\bold{r_\perp},\zeta)e^{k}_{J3}(\bold{r_\perp},\zeta)e^l_{J4}(\bold{r_\perp},\zeta) e^{i\Delta\kappa \zeta}}{\mathcal{N}_{J1}\mathcal{N}_{J2}\mathcal{N}_{J3}\mathcal{N}_{J4}},
    \label{eq:overlap_twodaggers_ring}
\end{align}
where $\Delta \kappa = \kappa_{J1} + \kappa_{J2} - \kappa_{J3} - \kappa_{J4}$. Unlike in Eq. \eqref{eq:overlap_twodaggers} for the waveguide, here the phase matching condition is contained in the effective area, since the mode profiles generally depend on the coordinate $\zeta$ along which the field propagates. The waveguide mode profiles $\bold{e}(x,y)$ do not depend on the direction of propagation, so the integral over $z$ and the $e^{i\Delta k z}$ term can be separated from the definition of the effective area. The phase matching condition then appears in the sinc terms in the nonlinear Hamiltonians for the waveguide, rather than in the effective area as we see in the ring system.

For processes involving three raising operators, we have 
\begin{align}
    \frac{e^{i\Phi_{J1,J2,J3,J4}}}{\mathcal{A}_{J1,J2,J3,J4}} =\frac{1}{\mathcal{L}} \frac{\int d\bold{r}_{\perp}d\zeta (\chi_{3}^{ijkl}(\bold{r}_{\perp})/\bar{\chi}_{3}) e^{*i}_{J1}(\bold{r_\perp},\zeta)e^{*j}_{J2}(\bold{r_\perp},\zeta)e^{*k}_{J3}(\bold{r_\perp},\zeta)e^l_{J4}(\bold{r_\perp},\zeta) e^{i\Delta\kappa \zeta}}{\mathcal{N}_{J1}\mathcal{N}_{J2}\mathcal{N}_{J3}\mathcal{N}_{J4}},
    \label{eq:overlap_threedaggers_ring}
\end{align}
with $\Delta \kappa = \kappa_{J1} + \kappa_{J2} + \kappa_{J3} - \kappa_{J4}$. 

\end{document}